\documentclass[manuscript,screen]{acmart}

\AtBeginDocument{%
  }

\setcopyright{none}
\copyrightyear{2018}
\acmYear{2018}
\acmDOI{XXXXXXX.XXXXXXX}

\acmConference[Conference acronym 'XX]{Make sure to enter the correct
  conference title from your rights confirmation emai}{June 03--05,
  2018}{Woodstock, NY}
\acmISBN{978-1-4503-XXXX-X/18/06}

\usepackage[ruled,vlined]{algorithm2e}

\usepackage[capitalize,noabbrev]{cleveref}

\begin{document}

\title{Uncovering Disparities in Rideshare Drivers’ Earning and Work Patterns: A Case Study of Chicago}




\author{Hy Dang}
\email{hdang@nd.edu}

\affiliation{%
  \institution{University of Notre Dame}
  \city{Notre Dame}
  \state{Indiana}
  \country{USA}
}

\author{Yuwen Lu}
\affiliation{%
  \institution{University of Notre Dame}
  \city{Notre Dame}
  \state{Indiana}
  \country{USA}
}

\author{Jason Spicer}
\affiliation{%
 \institution{City University of New York}
 \state{New York}
 \country{USA}}

\author{Tamara Kay}
\affiliation{%
  \institution{University of Notre Dame}
  \state{Indiana}
  \country{USA}
}
\author{Di Yang}
\affiliation{%
  \institution{Morgan State University}
  \city{Baltimore}
  \state{Maryland}
  \country{USA}}

\author{Yang Yang}
\affiliation{%
  \institution{University of Notre Dame}
  \city{Notre Dame}
  \state{Indiana}
  \country{USA}
}
\author{Jay Brockman}
\affiliation{%
  \institution{University of Notre Dame}
  \city{Notre Dame}
  \state{Indiana}
  \country{USA}
}
\author{Meng Jiang}
\affiliation{%
  \institution{University of Notre Dame}
  \city{Notre Dame}
  \state{Indiana}
  \country{USA}
}
\author{Toby Jia-Jun Li}
\affiliation{%
  \institution{University of Notre Dame}
  \city{Notre Dame}
  \state{Indiana}
  \country{USA}
}
\renewcommand{\shortauthors}{Dang et al.}

\begin{abstract}
Ride-sharing services are revolutionizing urban mobility while simultaneously raising significant concerns regarding fairness and driver equity. This study employs Chicago’s Trip Network Provider dataset to investigate disparities in ride-sharing earnings between 2018 and 2023. Our analysis reveals marked temporal shifts, including an earnings surge in early 2021 followed by fluctuations and a decline in inflation-adjusted income, as well as pronounced spatial disparities, with drivers in Central and airport regions earning substantially more than those in peripheral areas. Recognizing the limitations of trip-level data, we introduce a novel trip-driver assignment algorithm to reconstruct plausible daily work patterns, uncovering distinct driver clusters with varied earning profiles. Notably, drivers operating during late-evening and overnight hours secure higher per-trip and hourly rates, while emerging groups in low-demand regions face significant earnings deficits. Our findings call for more transparent pricing models and a re-examination of platform design to promote equitable driver outcomes.
\end{abstract}
\maketitle

\section{Introduction}

Ride-sharing services have become an important part of urban mobility, changing the ways people access transportation \cite{RAYLE2016168, RePEc:kap:transp:v:46:y:2019:i:6:d:10.1007_s11116-018-9923-2, NBERw22083}. Platforms such as Lyft, Uber, and DiDi serve as intermediaries that match passengers who request trips, to drivers who provide the ride. \cite{liu2023economic, banerjee2015pricing}. As of Q3 in 2024, Uber reportedly had 7.8 million drivers and couriers on the platform~\cite{uber_financials_2023}. 
Within this gig-based ecosystem, drivers operate as independent contractors, while ride-sharing platforms automatically assign trips for them to complete~\cite{de2024ridesourcing, cram2022examining, duggan2023algorithmic}. 

Despite their ubiquity, concerns over the fairness and equity in ride-sharing have been growing significantly \cite{liu2024evaluating, kumar2023using}. Studies highlight that these platforms' algorithms may inadvertently disadvantage racial, gender, and ethnic groups \cite{rovatsos2019landscape, nanda2020balancing}. For example, research has shown that African American passengers experience more frequent cancellations and longer wait times \cite{GE2020104205}. Policies around ride-sharing, such as California's Prop 22~\cite{california_prop22_2020}, stirred controversies around driver treatment. Drivers have been frequently expressing their dissent for unfair pay and mistreatment through frequent protests~\cite{bk_rideshare_2024, truthout_rideshare_2024}. These events raise pressing questions about how ride-sharing algorithms influence earnings and working conditions, and whether they reinforce or widen existing social inequities. Additionally, there have been research showing that algorithms in platforms not limited rideshare but other domains affects human behaviors and create biases and inequalities \cite{barocas2016big, rosenblat2016algorithmic, rosenblat2018uberland}.

However, systematically studying driver-level disparities remains challenging, mainly because of limited comprehensive public dataset \cite{chan2012ridesharing, allon2023impact} or the publicly released datasets from municipal governments \cite{chicago_tnp_2018, chicago_tnp_2023} remove all driver and passenger information for privacy. Researchers therefore lack the ability to link multiple trips to the same driver, making it impossible to observe real work patterns or total daily earnings.

In this paper, we propose and validate a case-study approach using Chicago’s Trip Network Provider dataset, one of the largest publicly available ride-share datasets. We begin by analyzing pricing factors and regional disparities in earning outcomes between 2018 and 2023, revealing time-dependent shifts in fare structures and persistent inequalities across different neighborhoods. Yet, we find that analyzing only trip-level summaries cannot fully depict the day-to-day driver experience. We therefore introduce an trip-driver assignment simulation algorithm that infers plausible driver workdays from otherwise anonymized trip data. By applying this framework to Chicago, we show that it is feasible to approximate drivers’ overall earning distributions and identify potential systematic disadvantages for certain driver groups, even without access to driver IDs or other sensitive platform data.

Our analysis reveals a surge in ride-sharing earnings in early 2021, followed by fluctuations and a \textbf{decline in inflation-adjusted income for drivers}. Spatially, a \textbf{widening gap in trip costs} emerges as Central and Airport regions consistently earn more than peripheral areas. Our trip assignment algorithm uncovers distinct driver groups with \textbf{varied work patterns and earning profiles}. Drivers who concentrate their work during late-evening or overnight hours consistently earn higher per-trip and hourly rates, even if they complete fewer trips compared to those driving predominantly during daytime. Additionally, spatial factors play a critical role---drivers focusing on premium areas like the airport or downtown earn substantially more than those serving peripheral neighborhoods, with a newly emerged group in the South Side in 2023, earning as low as \$12.74 per hour\footnote{Note that all we use the trip total, consisting of the trip fare, tip, and additional costs, to approximate driver earnings. The true net earning for drivers is lower but not available in public data.}. Overall, our results call for more transparent pricing models and a re-examination of platform design to foster equitable earning opportunities.



%
In summary, our contributions are threefold:
\begin{itemize}
\item We reveal disparities in Chicago’s ride-sharing trip data, showing significant shifts in trip allocations, earnings, and pricing factors since 2018.
\item 
We propose a simple trip-based driver-assigment simulation algorithm that generate potential driver-level activity from anonymized data, preserving privacy while able to observe how certain groups’ earnings diverge substantially from others.
\item 
We provide empirical evidence of equity gaps likely coming from algorithmic matching rules or regional fare structures, such findings that can inform more equitable policy and platform design.
\end{itemize}


\section{Related Work}

Research on ride-sourcing platforms has grown significantly, addressing issues from driver-passenger interactions and earnings disparity to the development of dispatching and matching algorithms \cite{chen2019value, ruch2020quantifying, hall2018analysis}. 
Below, we review key contributions in these areas and highlight how our work differs, particularly in terms of input data constraints and methodological goals.

\subsection{Ride-Sharing Driver Work Conditions in the Gig Economy}

Gig workers, particularly those in ride-sharing platforms like Uber and Lyft \cite{barrios2022launching}, face unique work conditions that have been extensively analyzed in recent studies. The flexibility of these platforms is often hailed as a major advantage, offering drivers the autonomy to choose their own hours~\cite{tan2021ethical}. However, this flexibility is accompanied by concerns about income instability, lack of benefits, and poor job security~\cite{rosenblat2018uberland}. Many ride-sharing drivers report feeling underpaid relative to their work hours, with some earning below minimum wage after accounting for expenses like fuel and vehicle maintenance~\cite{bajwa2018health, brown2024driving, mishel2018uber}. Additionally, the platform's algorithm-driven nature has been criticized for exacerbating power imbalances, as drivers have little control over fare pricing or working conditions~\cite{lobel2017gig, zhu2024gig, kloostra2022algorithmic}. Studies also highlight the impact of these conditions on drivers' well-being, including increased stress and burnout~\cite{berger2019uber}. These findings underscore the need for regulatory changes to improve working conditions in the gig economy, particularly for ride-sharing drivers.

\subsection{Algorithmic Fairness in Gig Work Platform}

A large body of work has examined socio-economic inequities and fairness concerns within the gig economy \cite{de2015rise, wood2019good, berg2015income}. For example, researchers analyzed the gender earnings gap among rideshare drivers, identifying disparities stemming from platform experience, geographic work preferences, and driving speed~\cite{cook2021gender}. Other lines of inquiry have revealed that these inequalities are not limited to gender factors alone. For instance, \cite{de2024ridesourcing} examined the relationship between economic inequality and the market share of ridesourcing services, showing that these platforms often capitalize on cost structures that can worsen income disparities.

Algorithmic management in gig work raises significant fairness concerns in task allocation, wage determination, and worker evaluation \cite{zhang2022algorithmic, duggan2020algorithmic, kadolkar2024algorithmic}. Workers contend with a severe information asymmetry, as platforms control crucial details about demand and algorithmic rules, leaving them to infer decision-making processes from online forums and peer discussions~\cite{rosenblat2016algorithmic}. Dynamic pricing mechanisms \cite{shapiro2020dynamic, van2020wage}, including surge pricing, often result in unpredictable and volatile earnings, causing both drivers and riders to view the system as exploitative~\cite{cramer2016disruptive}. Additionally, reputation systems that rely on customer ratings and automated metrics add to the stress, as a single poor rating can jeopardize a worker’s standing, despite factors beyond their control~\cite{lee2015working, rosenblat2018uberland}. This combination of ambiguous processes and automated decisions fosters an environment where gig workers feel the system is designed against them, highlighting serious ethical and fairness challenges \cite{wood2019good}.

These existing studies, while comprehensive, rely heavily on proprietary datasets containing detailed driver and passenger attributes, such as shift patterns, demographic information, and work behaviors. Such granular data supports fairness and inequality analyses but is not always publicly available. In contrast, our work utilizes only trip-level data, without access to underlying driver or passenger profiles.

\subsection{Ridesourcing Algorithms and Simulation Models}

In parallel, a substantial body of literature focuses on algorithmic strategies for matching drivers and passengers or for evaluating new policy interventions in ridesourcing. Previous studies have proposed optimization models and dispatching algorithms aimed at diverse objectives, such as maximizing platform revenue, minimizing passenger waiting times, or improving driver efficiency \cite{zhang2020pricing, schreieck2016matching, di2013optimization, cao2021optimization}. These methods typically integrate detailed driver and passenger information to simulate realistic markets, follows supply-demand dynamics, and test the impact of models under controlled conditions.

For instance, \cite{kucharski2022simulating} introduced a lightweight simulation model that matches supply (drivers) and demand (travelers) under various assumptions. Such frameworks are powerful for prototyping and testing policies, but they usually depend on inputs such as the number and locations of drivers, passenger arrival distributions, and driver acceptance behaviors. This reliance on detailed micro-level data and assumptions can limit the applicability of these models to large-scale, public datasets.

Additionally, there has been research conducted using public data. For example, \cite{nanda2020balancing} proposes NAdap, an algorithm aimed at balancing profit and fairness in ride-sharing systems during peak hours. Although the study utilized both real and synthetic datasets, the experiments were modeled and tested on a small scale, with the number of drivers ranging from approximately 50 for real-life datasets to 100 for synthetic datasets.

In contrast, our approach aims to work directly with large-scale, noisy, and inherently sparse public trip-level datasets—such as those provided by municipal transportation authorities. Without explicit driver or passenger information, or data on how trips were assigned or accepted, the challenge lies in designing simulation models that infer underlying dynamics and support analysis at scale. While prior simulation studies often rely on synthesized scenarios or small datasets, our method processes hundreds of thousands of real-world trips per day, sourced from publicly available data. To the best of our knowledge, this is the first effort to design simulation tools that operate solely on large-scale, trip-only data, enabling new forms of analysis to understand the inner workings of ridesourcing systems without needing sensitive or unavailable individual-level data.

In summary, while previous research has provided rich insights into fairness and has developed sophisticated ridesourcing algorithms, the reliance on proprietary or synthesized datasets often limits generalizability. Our contribution bridges this gap by proposing simulation frameworks that leverage publicly available trip-level data alone, offering a scalable and transparent means to study ridesourcing systems in real operational contexts.

\section{Methods}
\label{sec:methods}

Using the public Chicago ride-sharing dataset from 2018 to 2023~\cite{chicago_tnp_2018, chicago_tnp_2023}, our analysis focuses on temporal and regional ride price patterns, as well as driver work patterns. In all, we aim to answer the following research questions with our data analysis:

\textbf{RQ1:} How do ride-sharing trip \textit{costs change over time}?

\textbf{RQ2:} How do ride-sharing trip \textit{locations} impact trip costs?

\textbf{RQ3:} What are differences and similarities in ride-sharing \textit{drivers' work patterns}?


\subsection{Dataset}

The Chicago Transportation Network Providers Dataset includes all trips in Chicago reported by ride-sharing companies since 2018 to 2023. The dataset contains various data fields for each ride-sharing trip, such as start/end timestamps, trip durations and distances, locations (pickup and drop-off), and price-related information, which is further broken down into fares, tips, and fees. This large dataset contains around 300 million records from August 2018 to December 2022 and around 167 million records from January 2023 to the present\footnote{January 2025, as of writing this paper.}, averaging 200,000 to 300,000 records per day. It captures comprehensive aspects of ride-sharing trips and has been used in previous research to assess algorithm fairness~\cite{zhang2024data, manzo2022improving}.

Note that the earning numbers in our analysis were calculated using the fare and tips in the dataset. They do not include the portion taken by the platform, which is not available in the dataset. The last time Uber, the largest platform in Chicago, reported its ``take rate'' was in its 2023 earning report. Its average take rate was 29.3\% globally in the second quarter of 2023\footnote{\url{https://investor.uber.com/news-events/news/press-release-details/2023/Uber-Announces-Results-for-Second-Quarter-2023/default.aspx}}.  

\subsection{Data Analysis Metrics and Algorithms}
\subsubsection{Trip Price Shift Over Time (RQ1)}
To understand trip price temporal changes, we group all trips by months, and compare the \emph{average cost per hour} for each month's aggregated data\footnote{We use ``cost'' to refer to the trip total in the Chicago dataset, which includes fare, tip, and additional charges. Trip total is what a passenger pays the platform, thus we call it ``cost'' in our paper for ease of understanding.}. Our analysis of price temporal changes focus on the time period from November~2018 to December~2023.


\subsubsection{Trip Location Distribution and Spatial Pricing Patterns (RQ2)}
We also investigate how trips' pickup and drop-off locations are distributed across various regions in Chicago. Then, we compare the price of trips in different regions. We consolidate the trip data into seven regions (e.g., Central, North, West) following the classification used by the Chicago government\footnote{\url{https://data.cityofchicago.org/Community-Economic-Development/Boundaries-Planning-Regions-Map/2wek-zf5g}}. We aggregate the Chicago dataset by years to investigate these regional patterns, and left out 2018 since the dataset only contains 2 months for this year. Similar to above, we used average cost per hour as our metric to compare regional trip prices across years. 




\subsubsection{Driver-Level Work Pattern Analysis (RQ3)}
\label{sec:methods-driver-simulation}

A main goal of our data analysis is to understand ride-sharing drivers' earnings and work patterns. However, the Chicago ride-sharing dataset, similar to many other datasets in this domain, does not associate trips with either drivers or passengers. Such practice is to protect passengers' and drivers' privacy---once we link together all rides requested by a passenger or provided by a driver throughout the years, it is not impossible to uncover the person's identity from the activity patterns, even if the data is anonymized. As a result, from the Chicago dataset, it is difficult to identify driver profiles, posing a challenges for our analysis. This prevents us from estimating drivers' total earnings and work hours, which is necessary for studying biases, fairness among driver groups.

To address this challenge, we developed a trip-based driver assignment algorithm (\cref{alg:concise_trip_assignment}), which \emph{assigns} ride-sharing trips to hypothetical drivers under realistic spatiotemporal constraints. Our algorithm works under one assumption---if a driver takes a new trip after finishing a previous one, it is likely that the second trip's pickup location is close to the first one's drop-off location (\textit{spatial constraint}), and the second trip starts shortly after the first one (\textit{temporal constraint}). Based on this assumption, our algorithm assigns trips that are \textit{spatially} and \textit{temporally} close to each other to one hypothetical driver. We also make the assumption that a driver does not work continuously for more than 8 hours or 25 trips without a break.

\begin{algorithm}[ht]
\caption{Trip Assignment Simulation Algorithm}
\label{alg:concise_trip_assignment}
\DontPrintSemicolon

\KwIn{Data frame \(\mathit{df}\) with trips; time threshold \(\alpha\); distance threshold \(\textit{maxDist}\).}
\KwOut{Mapping of trips to driver routes.}

\While{\(\mathit{df}\) is not empty}{
  \(\textit{currentDriverTrips} \gets \{\}\); 
  \(\textit{currentTrip} \gets\) Randomly pick from \(\mathit{df}\)\;
  \(\textit{currentDriverTrips} \leftarrow \textit{currentTrip}\); 
  Remove \(\textit{currentTrip}\) from \(\mathit{df}\)\;

  \While{driver constraints not exceeded}{
    \(\mathit{candidates} \gets\) trips in \(\mathit{df}\) where 
      start time is in \([\textit{currentTrip.endTime}, \textit{currentTrip.endTime} + \alpha]\)\;
    Filter out candidates with distance to \(\textit{currentTrip.dropoff}\) 
      \(> \textit{maxDist}\)\;
    \If{\(\mathit{candidates}\) is empty}{\textbf{break}}
    \(\textit{nextTrip} \gets\) Randomly pick from top feasible candidates\;
    \(\textit{currentDriverTrips} \leftarrow \textit{nextTrip}\); 
    Remove \(\textit{nextTrip}\) from \(\mathit{df}\)\;
    \(\textit{currentTrip} \gets \textit{nextTrip}\);
  }

  Assign \(\textit{currentDriverTrips}\) to a new driver\;
}
\end{algorithm}

Algorithm~\ref{alg:concise_trip_assignment} describe the detailed procedure to assign trips to drivers. In the beginning, we randomly pick one unassigned trip as the starting point. Then, from other unassigned trips, we find trips that are spatiotemporally close to the selected trip, then append the new trip to the selected one. We iteratively apply this step to form a \emph{driver route}. When appending a new trip to a currently selected trip, we apply the following constraints:
\begin{enumerate}
    \item \textbf{Temporal constraint}: The new trip's start time must be after the current trip's end time but within a threshold $\alpha$ (e.g., 15 minutes).
    \item \textbf{Spatial constraint}: The distance between the current trip's drop-off location and the next trip's pickup location must be less than a specified threshold \texttt{max\_distance}.
    \item \textbf{Driver constraints}: Each driver is assumed to work for no more than 8 hours per day or a maximum of 25 trips, whichever is reached first.
\end{enumerate}

When there are multiple candidate trips that meet these constraints, we randomly pick one from the candidates. When running this algorithm on the Chicago dataset, we set the time threshold $\alpha = 0.25$ (hour) and maximum distance \text{maxDist} = 1 (mile). This results in a list of hypothetical drivers and their associated trips. 

After we assign trips to the hypothetical drivers, we used the KMeans algorithm~\cite{lloyd1982least} to cluster drivers into groups and analyze their work patterns. The drivers are clustered based on their associated trip features, including fares, earnings, pickup and drop-off locations, trip timestamps and distances. Drop-off and pickup timestamps are converted into three-hour intervals to improve clustering quality. 

Given the large size of the Chicago dataset, it is very computationally expensive to run the algorithm on all available data. Thus, we randomly select the second week of August in years 2019 (August 5-11) and 2023 (August 7-13) to conduct our analysis. To determine the optimal number of clusters, we compute the Silhouette Score~\cite{rousseeuw1987silhouettes} for a range of cluster values from 4 to 16 and select the value with the highest score as the final number of clusters. As a result, we had 9 driver groups for the week in 2019 and 11 groups for 2023.

After clustering, for each driver group, we evaluate their work pattern on metrics including frequency of pickup/drop-off trips in community areas, work time periods, and aggregated daily trip earnings\footnote{Note that the Chicago dataset does not contain driver earning information but just trip cost for passengers. We use the trip total as a reasonable proxy to estimate driver earning in analysis of hypothetical driver earnings.}. Since the dataset lacks detailed wait-time information, we apply a 0.25-hour wait time to all trips to approximate the real work hours.

\subsubsection{Summary}
Using these three main types of analysis metrics, we provide a comprehensive understanding of the ride-sharing landscape in Chicago. By examining temporal shifts in trip costs (RQ1), we assess how ride prices have evolved over time and whether these changes align with broader economic trends. The spatial pricing analysis (RQ2) enables us to evaluate regional variations in ride costs. Finally, through our driver work pattern analysis (RQ3), we approximate driver activity and earnings despite the dataset's anonymization, allowing us to cluster and distinguish different driver behaviors. Together, these analyses offer a multi-faceted view of ride-sharing operations, helping us uncover key trends and potential disparities in the platform algorithms.
\section{Results}
We structure our results to directly address each Research Question (RQ), integrating findings from both the raw Chicago dataset (RQ1, RQ2) and after applying our driver assignment algorithm (RQ3). 

\subsection{RQ1: How do ride-sharing trip costs change over time?}
\label{RQ1}
\begin{figure}[h]
  \centering
  \includegraphics[width=0.9\linewidth]{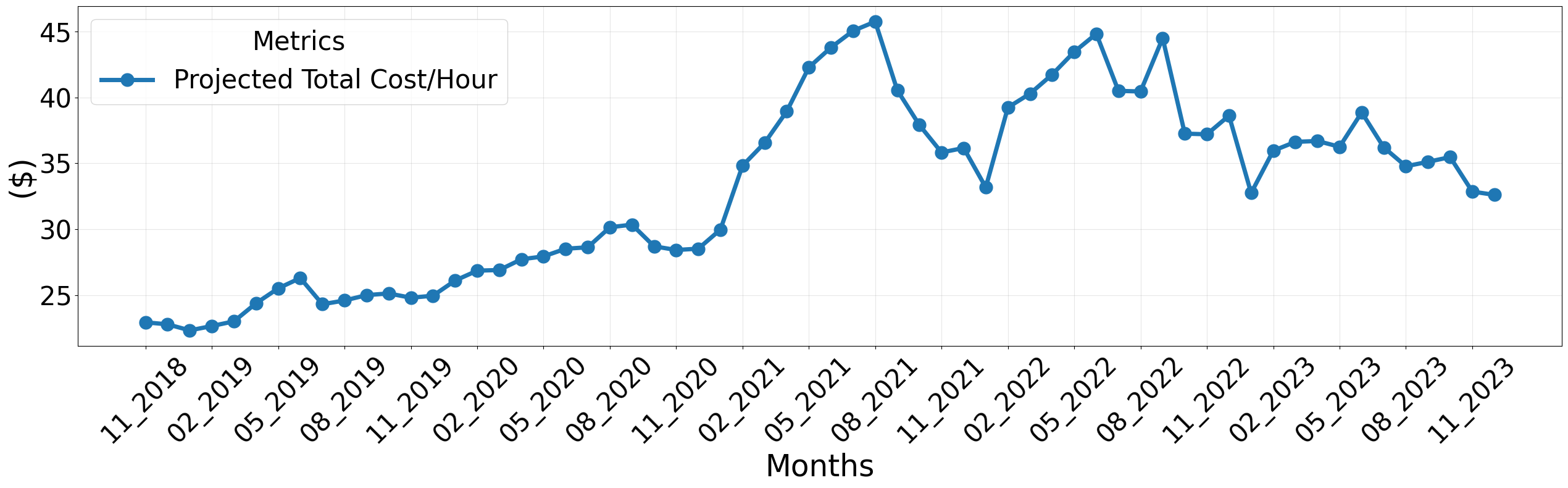}
  \caption{Cost per driving hour trends from November 2018 to November 2023. A significant increase in earnings per trip and projected total earnings per driving hour begins in January 2021, peaking in August 2021. After this peak, all metrics fluctuate and gradually decline throughout 2023, with no major increases observed since late 2021.}
  \label{by_hour}
\end{figure}



\label{sec:results-pricing-stablization}

\paragraph{\textbf{General Trends in Pricing Over Time.}}
Figure~\ref{by_hour} illustrates the projected total cost per driving hour from November 2018 to November 2023. The data shows a \textbf{gradual increase in earnings per hour from late 2018 to early 2020}, followed by a steep surge beginning in early 2021, peaking around August 2021. After this peak, there is a notable decline in late 2021, followed by fluctuations and a secondary surge until mid-2022. From mid-2022 onwards, the overall trend declines, with no major increases observed through the end of 2023.

These patterns suggest that platform pricing strategies have potentially undergone multiple shifts, particularly in response to broader economic and market conditions.

\paragraph{\textbf{Stagnation in Driver Earnings vs. Inflation-Adjusted Wages.}}

Although hourly costs remained relatively stable from late 2021 onward, they did not keep pace with inflation. According to the Bureau of Labor Statistics, the Consumer Price Index for All Urban Consumers (CPI-U) in the Chicagoland area rose by approximately 15\% between 2021 and 2023~\cite{BLS_CPI}. However, Figure~\ref{by_hour} suggests that driver earnings did not experience a corresponding increase.

This stagnation raises serious equity and long-term sustainability concerns. Even if nominal fares remain steady, the inflation-adjusted value of driver earnings is decreasing. As a result, many drivers may be experiencing a decline in purchasing power, making gig work less financially viable over time. The lack of alignment between earnings and inflation suggests that ride-sharing platforms may not be adjusting pay rates in response to economic conditions, raising concerns about fair wage structures and transparency~\cite{angrist2021uber}.



\subsection{RQ2: How do ride-sharing trip locations impact trip costs?}

To examine regional trip cost differences, we conducted a analysis of trip volumes and earnings in 7 main regions Chicago from 2018 to 2023. Here, we highlight findings from 2019 and 2023. Additional figures for other years are included in Appendix~\ref{sec:appendix}.

\paragraph{\textbf{Spatial Demand Shift vs. Persistent Regional Disparities.}}

\begin{figure}[ht!]
  \centering
  \includegraphics[width=0.95\linewidth]{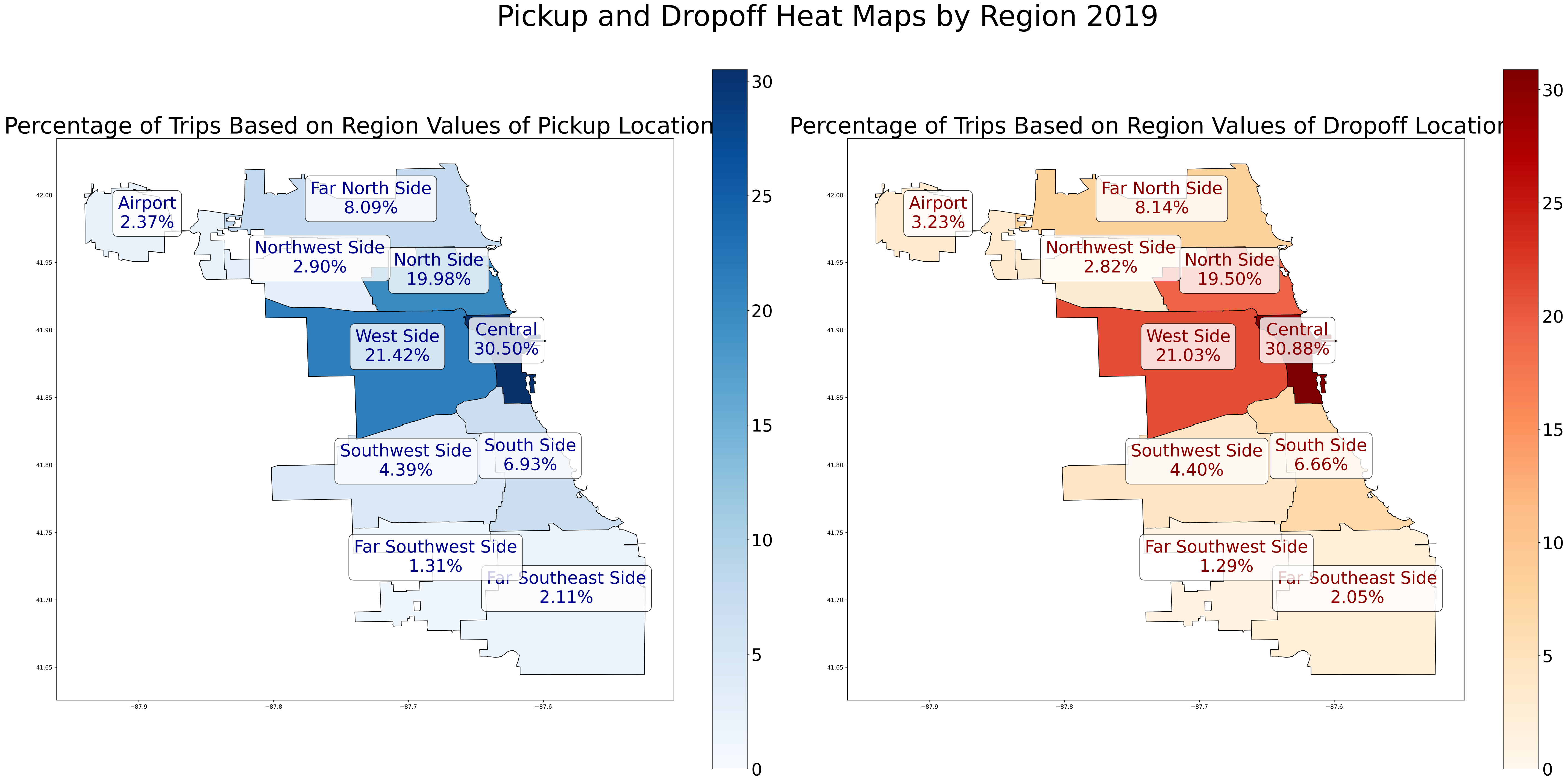}
\caption{Heat maps of pickup and dropoff trip distributions across Chicago in 2019. The highest concentration of activity was in the Central region, accounting for 30.50\% of pickups and 30.88\% of dropoffs, followed by the West Side. Other regions, such as the South Side and Far Southeast Side, shows lower levels of ride-hailing activity, with the Far Southwest Side being the least active area, contributing only 1.31\% of pickups and 1.29\% of dropoffs.}
    \label{fig:distribution_2019}
\end{figure}

\begin{figure}[H]
  \centering
  \includegraphics[width=0.95\linewidth]{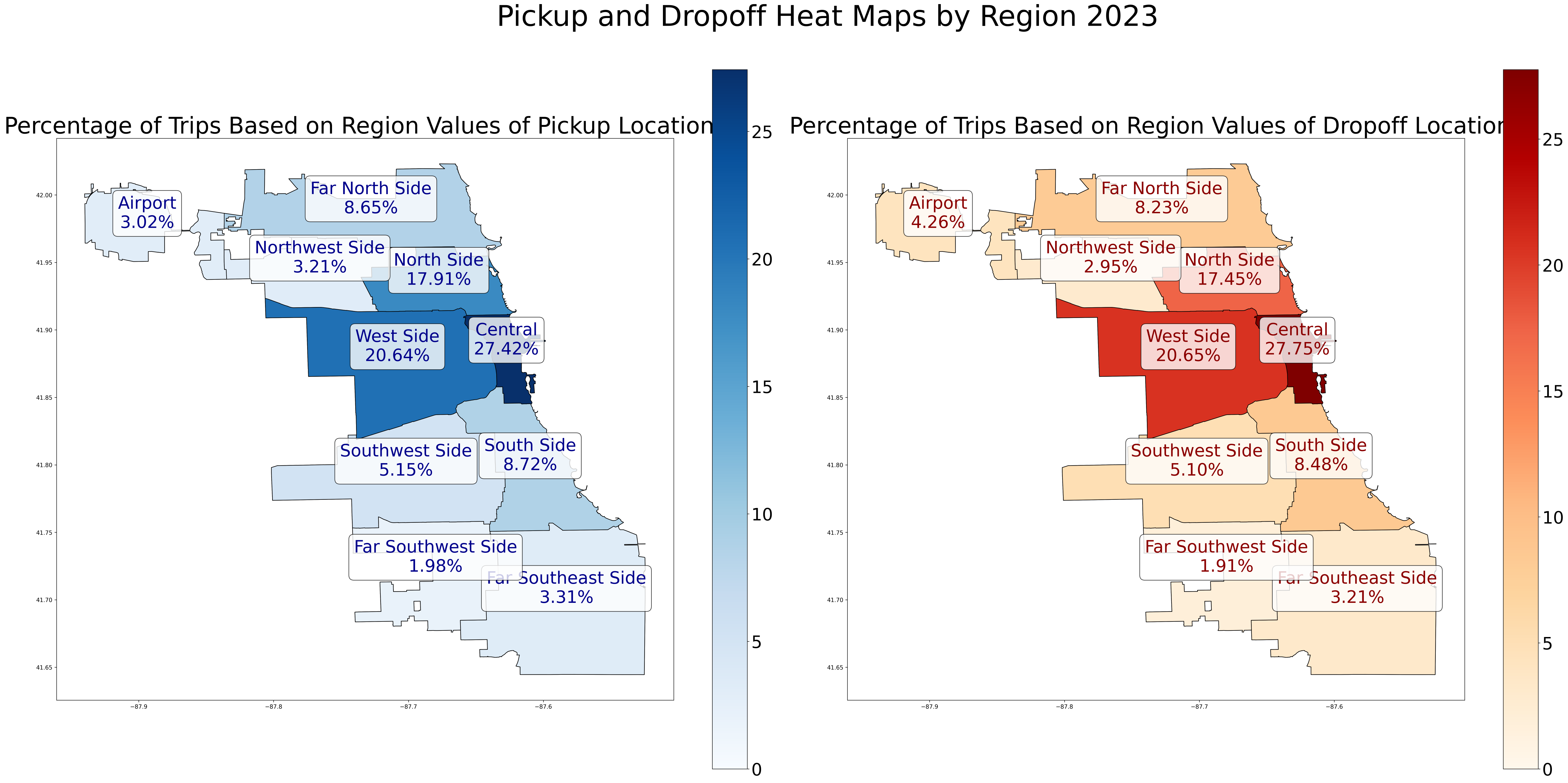}
\caption{Heat maps of pickup and dropoff trip distributions across Chicago in 2023. The Central region remains the most active, with 27.42\% of pickups and 27.75\% of dropoffs, followed by the West Side. The South Side and Far Southeast Side show increased activity compared to 2019, the Far Southwest Side continues to account for the lowest share, with 1.98\% of pickups and 1.91\% of dropoffs.}
    \label{fig:distribution_2023}
\end{figure}

Figures \ref{fig:distribution_2019} and \ref{fig:distribution_2023} compare the percentage of pickup and drop-off trips across major Chicago regions in 2019 and 2023. The Central region held the highest share of rides in both years---about 30\% in 2019 and nearly 27\% in 2023---indicating that downtown and nearby neighborhoods remain a focal point for ride-hailing. However, from 2019 to 2023, there was a \textbf{notable increase} in the share of trips in the South Side and Far Southeast Side, suggesting a gradual dispersal of ride-hailing activity beyond traditional high-density zones such as the Central or West Side. These findings also inform our subsequent simulation analysis, in which we explore how pricing and driver relocation strategies align with observed real-world changes in spatial demand.

\paragraph{\textbf{Emergence of Low-Cost Zones in 2023.}}

\begin{figure}[H]
  \centering
  \includegraphics[width=0.95\linewidth]{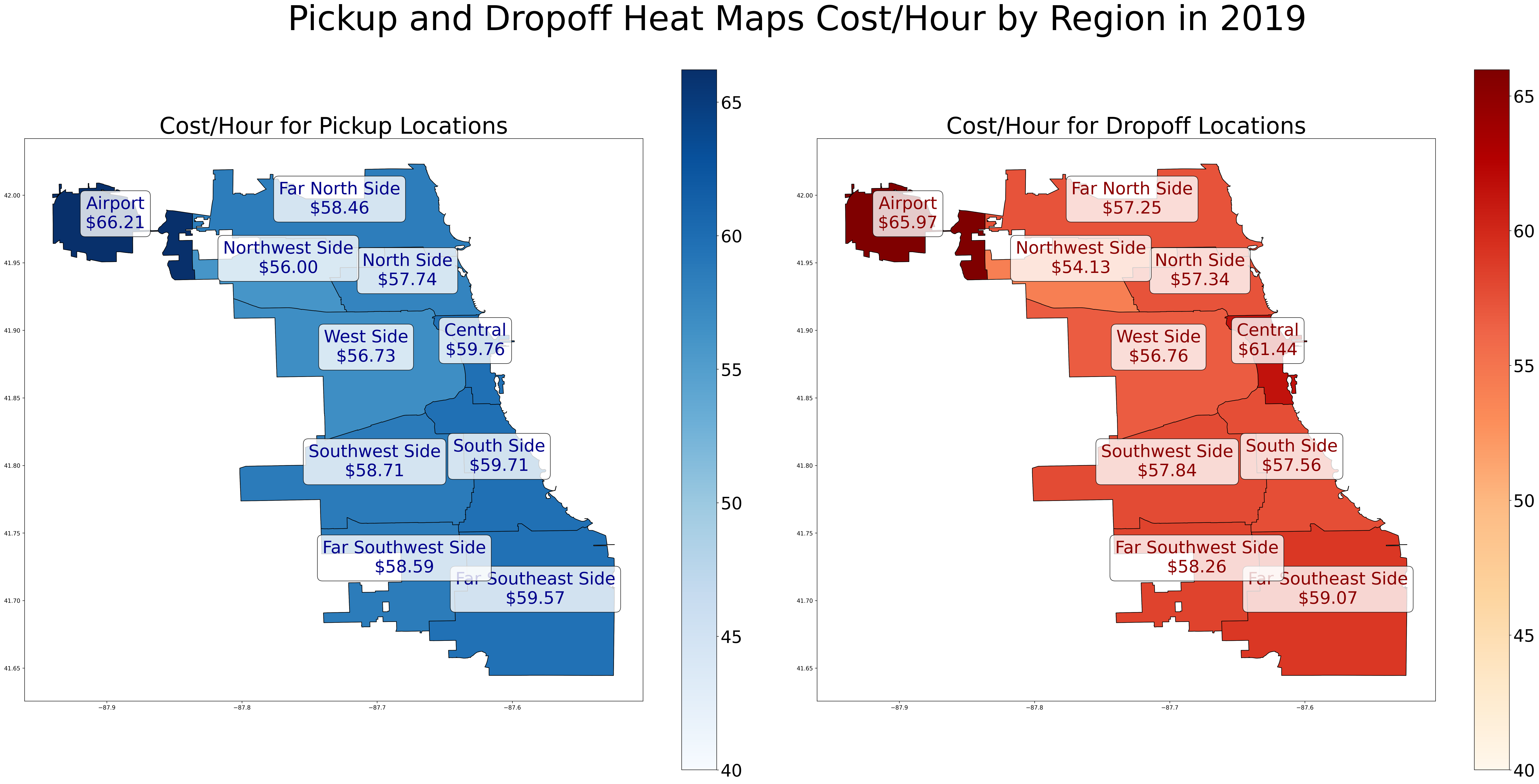}
\caption{Projected hourly driving costs in 2019. The highest cost area is in the Airport region, with \$66.21/hour for pickups and \$65.97/hour for dropoffs. The Central and Southwest Side regions with the top earnings regions. In contrast, the Northwest Side recorded the lowest earnings, with \$56/hour and \$54.13/hour for pickup and dropoff, respectively.}
    \label{earnings_2019}
\end{figure}

\begin{figure}[H]
  \centering
  \includegraphics[width=0.95\linewidth]{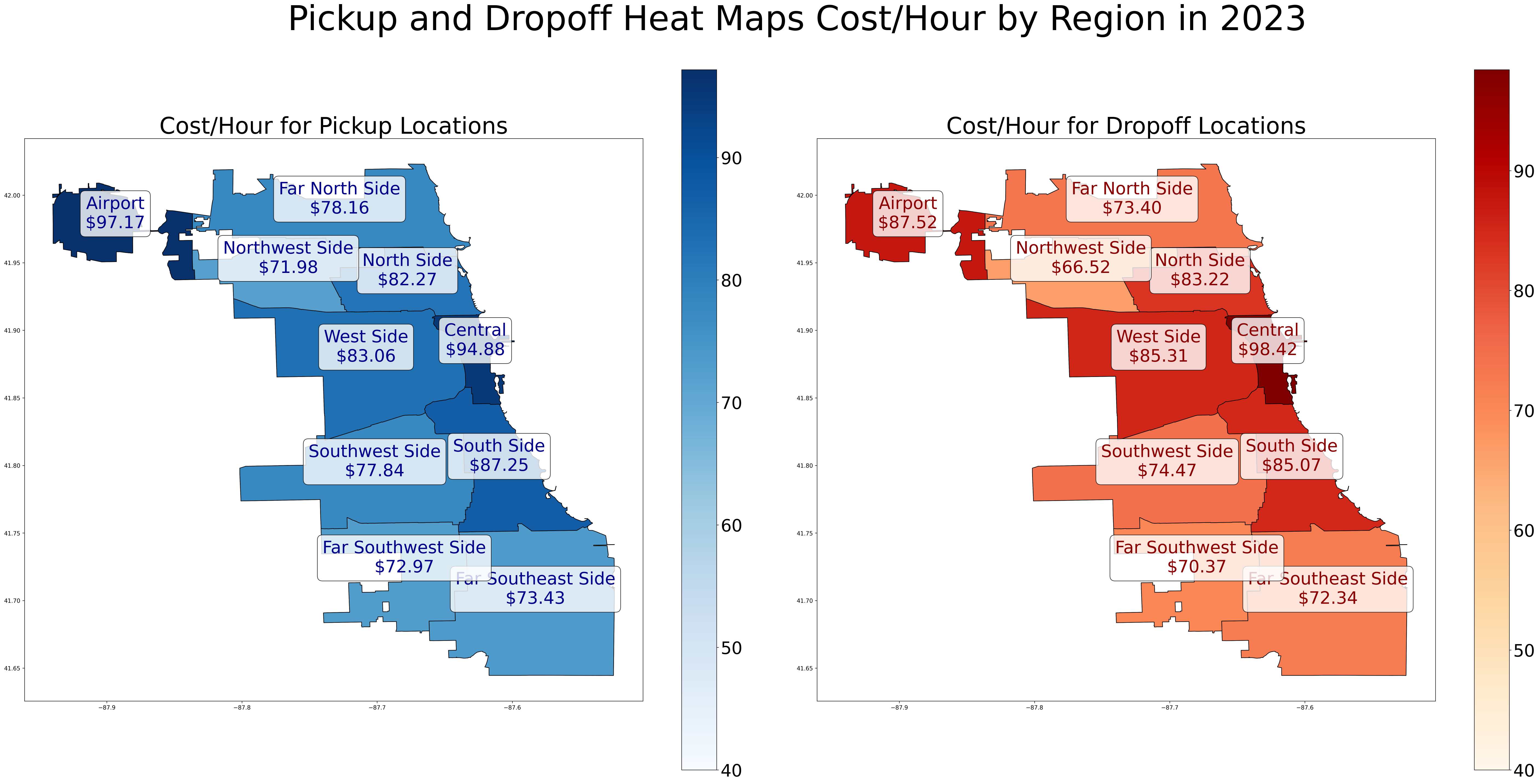}
\caption{Average hourly trip costs in 2023. The highest cost remains in the Airport region, with \$97.17/hour for pickups and \$87.52/hour for dropoffs. The Central and Southwest Side regions are among the top earning areas. In contrast, the Far Southeast Side and Far Southwest Side recorded the lowest earnings, with \$30.06/hour and \$31.14/hour for pickup and dropoff, respectively.}
    \label{earnings_2023}
\end{figure}

Figures~\ref{earnings_2019} and~\ref{earnings_2023} show that although nearly every region experienced cost or price growth---for example, the Central region increased from about \$60/hour to over \$95/driving hour, and the Southwest Side rose from \$58/hour to \$75/hour---areas in the far south nonetheless trail considerably behind both the Airport and downtown core. We hypothesize that these upward trends partially stem from the pricing-model adjustments discussed in Section \ref{RQ1}.

Despite the gains, disparities among non-airport regions have \textbf{widened significantly}. For instance, while the Central outperformed many other neighborhoods in 2019, with hourly driving cost of \$59.76--\$61.44, its advantage over the low-cost region (Northwest Side, with hourly driving cost of \$56--\$54.13) was relatively modest (7--13\%). By 2023, however, the gap between top-earning regions (e.g., \$94.88--\$98.42 in the Central region) and low-cost areas like the Far Southwest Side (\$72.97--\$70.37) has reached to 30--40\%. In particular, the South Side, Far Southwest Side, and Far Southeast Side now show substantially lower costs than other regions---a distinction that was far less pronounced in 2019---underscoring the increasing income inequality across Chicago’s ride-hailing landscape.

This \textbf{increasing income inequality} suggests that pricing model adjustments are not benefiting all drivers equally, reinforcing disparities among different Chicago neighborhoods.


\paragraph{\textbf{Airport Premium vs. Outlying Gaps.}}
The airport region exhibits the highest trip cost growth, increasing from approximately \$66.21--\$65.97 per hour in 2019 to \$97.17--\$87.52 per driving hour in 2023. Although the airport region consistently shows much higher trip costs, we hypothesize that this might be partially explained by its geographic context---longer trips typically originate in other regions---and by extended wait times commonly encountered at the airport. 
Overall, despite relatively stable trip distributions, regional earnings have shifted considerably. This emphasizes the role of location-specific demand and fare policies in shaping driver incomes over time, a pattern we also observe in our simulation models.


\subsection{RQ3: How do ride-sharing drivers' work patterns look like?}
\begin{table}[ht]
\centering
\caption{
  Earning Metrics for Simulated Driver Clusters (2nd week of August in 2019 and 2023). In 2019, Cluster 0 had the highest total income (\$288.70) and fares (\$216.10) where they perform the most number of trips, while Cluster 6 and Cluser 8 led in earnings per driving hour (\$57.89) and earnings per hour (\$29.77), respectively. Cluster 3 earned the least in hourly earnings at \$48.60. By 2023, incomes increased overall, with Cluster 9 achieving the highest earnings per driving hour (\$96.66) and per trip (\$21.28). Cluster 8, 10 had the lowest per-trip earnings (\$12.74) and total income (\$87.82), respectively, reflecting shifts in income distribution over time.
  \newline
  \emph{Definitions:} 
  E/Trip = average earning \emph{per trip}; 
  E/DriveHr = average earning \emph{per driving hour}; 
  Est.\ E/Hr(+0.25) = average estimated earning \emph{per hour including a 15-min wait assumption}; 
  Total Fares = total amount of Fares \emph{per date from trips}; 
  Total Inc. = total income \emph{per date after combining all fares, fees, and tips from trips.} 
  \newline
  \emph{Note: Bolded values in each column represent the highest value; underlined values represent the lowest. It is important to note that driver groups in the two timeframes are not directly related.}
}
\label{tab:predict_cluster_case_2_no_cluster_9}
\begin{tabular}{ccccccc}
\toprule
\textbf{Driver} & 
\textbf{\#Trips} & 
\textbf{E/Trip (\$)} & 
\textbf{E/DriveHr (\$)} & 
\textbf{Est.\ E/Hr (+0.25) (\$)} & 
\textbf{Total Fares (\$)} & 
\textbf{Total Inc.\ (\$)}\\
\midrule
\multicolumn{7}{c}{\textbf{Simulated Driver Groups in 2019}}\\
\midrule
0 & \textbf{20.91} & 13.23 & 48.05 & 25.70 & \textbf{216.10} & \textbf{288.70} \\
1 & 7.48 & \underline{11.58} & 51.40 & 24.89 & 66.68 & 90.17 \\
2 & \underline{4.49} & 12.33 & 48.60 & 25.00 & \underline{43.23} & \underline{57.85} \\
3 & 6.30 & 12.43 & 43.94 & \underline{23.75} & 61.16 & 81.41 \\
4 & 5.53 & 12.14 & 47.71 & 24.53 & 51.43 & 69.80 \\
5 & 5.63 & 13.22 & 49.22 & 26.04 & 58.28 & 77.80 \\
6 & 5.30 & 11.66 & \textbf{57.89} & 26.44 & 48.44 & 64.49 \\
7 & 6.47 & 13.22 & 46.81 & 25.33 & 66.83 & 89.27 \\
8 & 5.52 & \textbf{13.95} & \underline{40.53} & \textbf{29.77} & 60.93 & 80.92 \\
\midrule
\multicolumn{7}{c}{\textbf{Simulated Driver Groups in 2023}}\\
\midrule
0 & \textbf{20.87} & 17.08 & 62.59 & 33.44 & \textbf{277.59} & \textbf{374.54} \\
1 & 7.97 & 15.52 & 69.37 & 33.52 & 97.63 & 129.15 \\
2 & 6.77 & 17.16 & 59.81 & 32.60 & 91.12 & 121.25 \\
3 & 5.88 & 18.75 & 70.41 & 37.17 & 84.82 & 115.73 \\
4 & \underline{4.57} & 18.68 & 76.01 & 38.58 & \underline{65.70} & 89.55 \\
5 & 5.48 & 17.20 & 68.31 & 34.96 & 71.71 & 98.16 \\
6 & 5.34 & 17.38 & \textbf{88.12} & \textbf{39.73} & 76.50 & 96.59 \\
7 & 6.85 & 17.40 & 60.80 & 33.19 & 92.64 & 125.10 \\
8 & 6.51 & \underline{12.74} & \underline{52.47} & \underline{26.76} & 68.96 & 88.91 \\
9 & 5.31 & \textbf{21.28} & \textbf{96.66} & 46.40 & 90.20 & 118.42 \\
10 & 5.32 & 15.68 & 58.39 & 31.00 & 66.67 & \underline{87.82} \\
\bottomrule
\label{table:earning_driver}
\end{tabular}
\end{table}

Using our trip assignment simulation algorithm (\cref{alg:concise_trip_assignment}), we analyzed driver earnings and trip behavior for the second week of August in 2019 and 2023.

\paragraph{\textbf{Clustering Reveals Distinct Earning Profiles.}}

Our simulation for the two weeks in 2019 and 2023 yielded a total of 364,452 and 259,812 drivers, respectively. To focus on active drivers, we filtered for those completing more than two trips daily, resulting in samples of 164,234 drivers in 2019 and 114,920 in 2023. Using KMeans clustering with Silhouette Score optimization, we identified 9 distinct driver groups in 2019 and 11 in 2023. These clusters revealed diverse patterns in trip frequencies, average fares, and hourly earnings, demonstrating significant variations across different segments of the driver population (\cref{tab:predict_cluster_case_2_no_cluster_9}). The t-SNE visualizations in \cref{sec:appendix} illustrate the spatial distribution of these cluster groups, validating both our simulation methodology and clustering approach.

Table~\ref{tab:predict_cluster_case_2_no_cluster_9} presents some key findings regarding the identified driver profiles:

\begin{enumerate}
    \item In 2019, Cluster 0 had the highest total income (\$288.70) and fares (\$216.10), as they performed the most trips, while Cluster 7 achieved the highest earnings per hour (\$60.53).
    \item By 2023, overall incomes increased, with Cluster 9 achieving the highest earnings per drive hour (\$96.66) and per trip (\$21.28).
    \item However, disparities widened, with Cluster 8 and Cluster 10 earning significantly less (\$12.74 per trip and \$87.82 total income, respectively).
\end{enumerate}

These results may suggest that driver work strategies have become more differentiated, with some groups benefiting from shifts in demand while others are increasingly disadvantaged.

\paragraph{\textbf{Emerging New Driver Groups.}}

Our clustering approach also revealed new driver groups in 2023 that were not present in 2019, particularly those concentrating a higher proportion of their trips in the South Side and Far Southeast Side (Driver Group 8) and the Northwest Side (Driver Group 10). This suggests emerging spatial specialization, where certain drivers may be systematically targeting or being assigned lower-demand neighborhoods. The growing concentration of drivers in historically lower-paying areas raises questions about platform steering and whether algorithmic matching is reinforcing existing income disparities.


\paragraph{\textbf{Differences in Driver Earnings Based on Temporal Patterns.}}

\begin{figure}[ht]
  \centering
  \includegraphics[width=\linewidth]{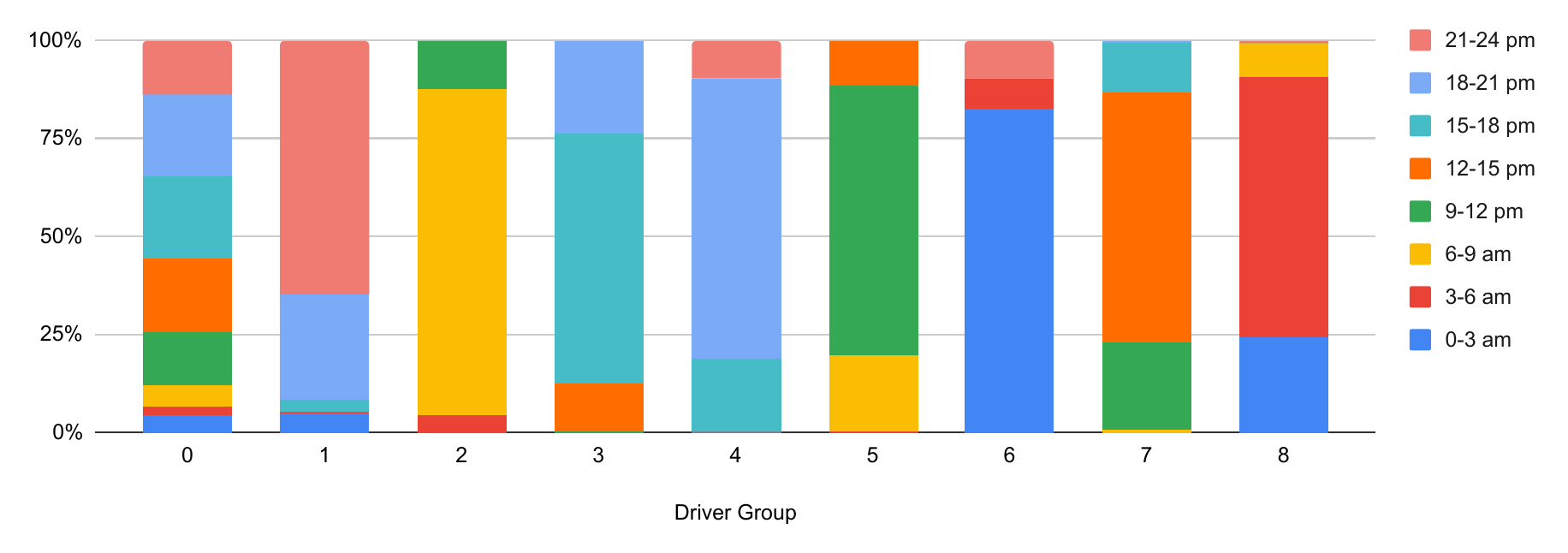}
  \caption{Proportion of Trip in Different Time Intervals for Different Driver Group Clusters in the Week of 2019-08-05. The proportion of trips for different driver group clusters in 2019 shows distinct patterns across various time intervals. Certain driver groups concentrate their activity in specific periods, such as late evening (with more than 60\% of trips starting within 21--24 hours for Driver Group 1) or early morning (with about 80\% of trips starting within 0--3 hours for Driver Group 6). Meanwhile, Driver Group 0 have a more evenly distributed trip pattern throughout the day.
}
  \label{figure:time_chunk_2019}
\end{figure}

\begin{figure}[H]
  \centering
  \includegraphics[width=\linewidth]{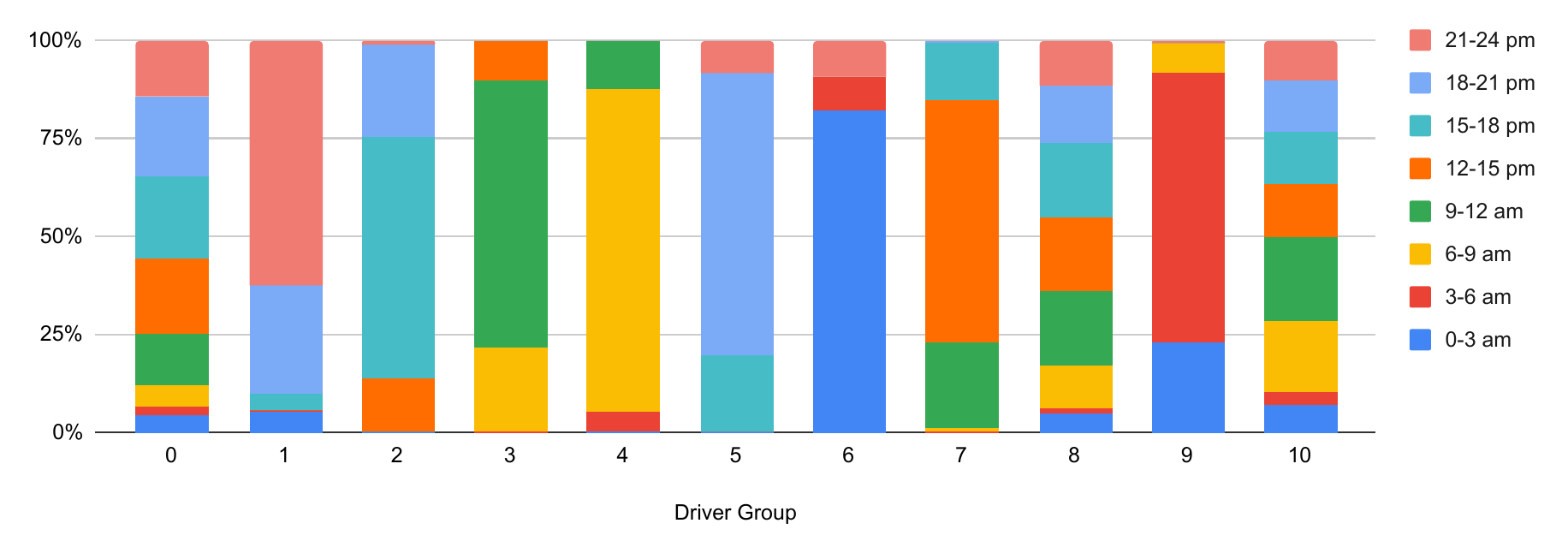}
  \caption{Proportion of Trip in Different Time Interval For Different Driver Group Cluster in the week of 2023-08-07. The proportion of trips for different driver group clusters in 2023 demonstrates a similar variety of temporal driving patterns as in 2019. Most of the groups focus heavily on specific intervals, while others distribute their trips more broadly across the day (Driver Group 0, 8 and 10).}

  \label{figure:time_chunk_2023}
\end{figure}

Figures~\ref{figure:time_chunk_2019} and~\ref{figure:time_chunk_2023} illustrate how different driver groups distribute their trips across the day, showing strong variations in work schedules and earnings potential. In both years, certain groups concentrate on \textbf{late-evening} or \textbf{overnight} windows and \textbf{achieve relatively high per-trip revenue}, while others who maintain heavier daytime schedules do not necessarily attain higher hourly wages. For instance, one group logs the greatest number of trips between 9AM and 6PM (Driver 0 in both 2019 and 2023) but still reports moderate or even below-average hourly earnings. Meanwhile, a different group working fewer trips between 9PM and 12AM records top-tier rates (Driver 1 in 2019). Similarly, two groups driving the same time blocks can yield markedly different pay if they serve distinct neighborhoods or exploit surge-pricing periods differently. Overall, these patterns highlight that although temporal availability matters, it intersects with factors like spatial demand, platform matching, and each driver’s individual strategies, resulting in significant earning disparities among those who choose similar—or even overlapping—hours. 

\begin{figure}[H]
  \centering
  \includegraphics[width=0.85\linewidth]{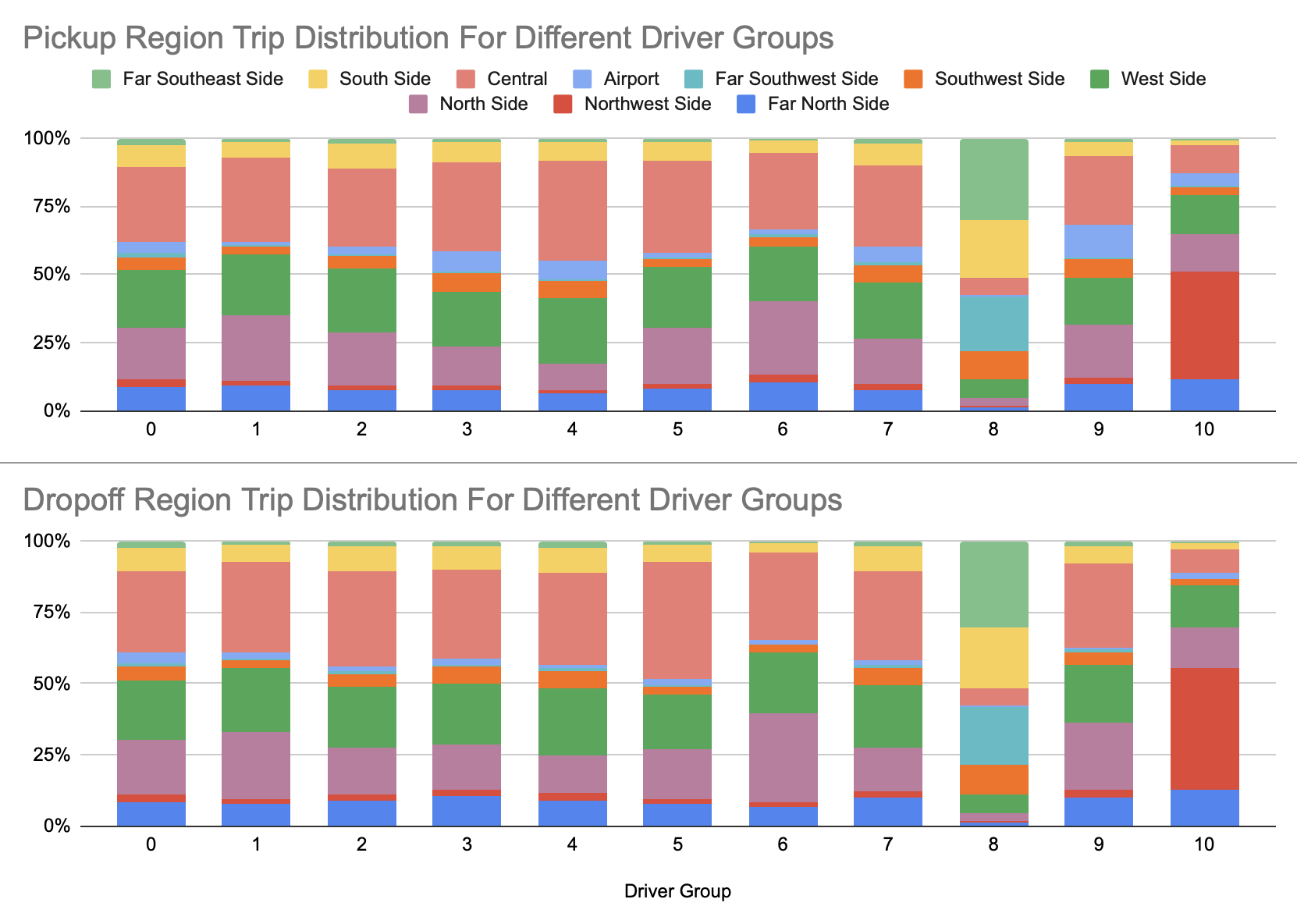}
\caption{Proportion of Trips by Pickup and Dropoff Locations for Different Driver Group Clusters in the Week of 2023-08-07. The Central region remains the most active area for most driver groups (except Driver Group 8 and 10). Driver Group 8 has also emerged as a distinct cluster focusing on low-activity areas, with about 30\% and 20\% of their trips occurring in the Far Southeast Side and South Side, respectively. Additionally, Driver Group 10 now concentrates more than 40\% of their trips in the Northwest Side, a pattern not observed in 2019.}

  \label{figure:dropoff_pickup_2023}
\end{figure}

\paragraph{\textbf{Differences in Driver Earnings Based on Regional Distribution.}}
\begin{figure}[ht!]
  \centering
  \includegraphics[width=0.85\linewidth]{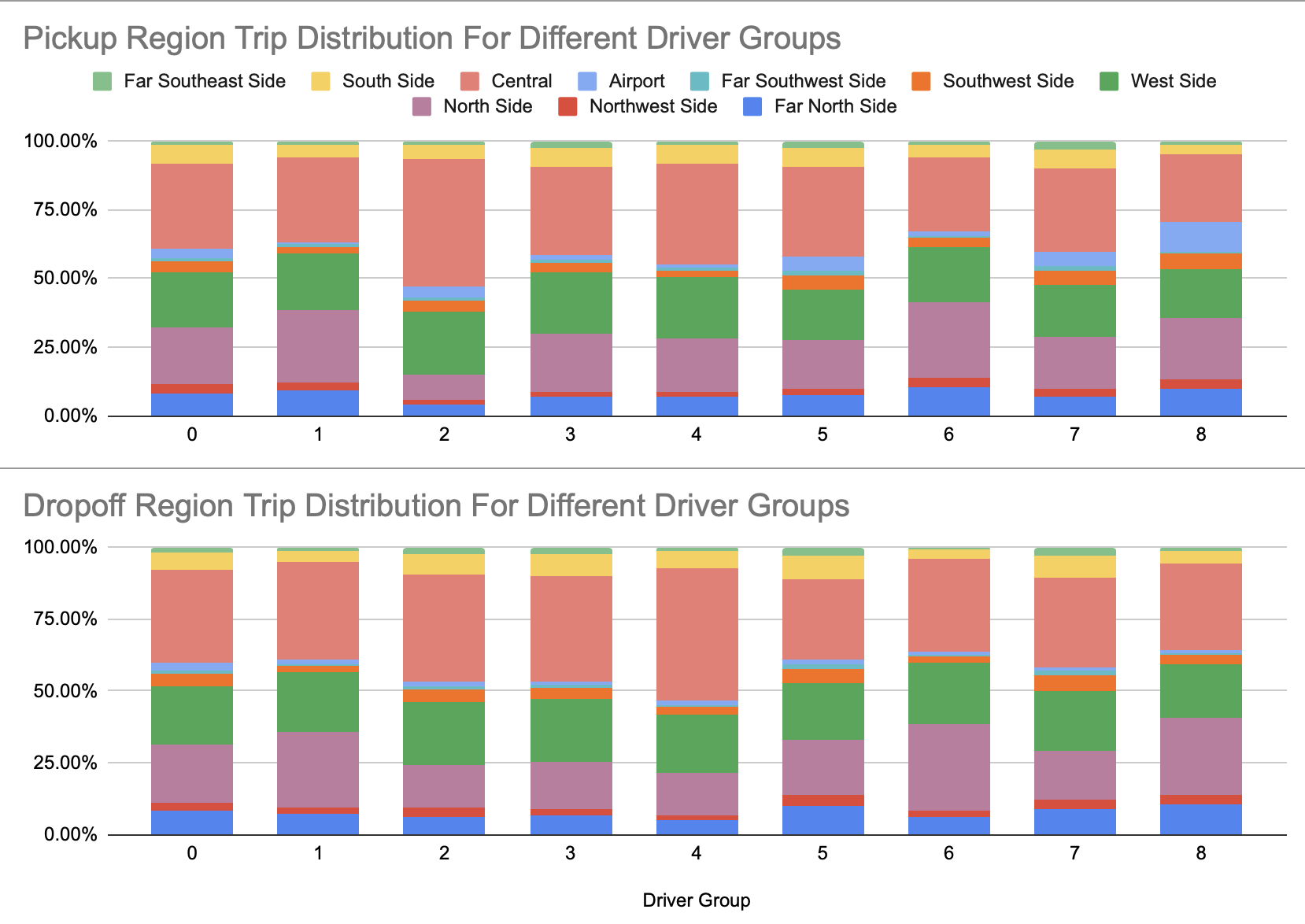}
\caption{Proportion of Trips by Pickup and Dropoff Locations for Different Driver Group Clusters in the Week of 2019-08-05. The Central region shows the highest concentration of trips across most driver groups, accounting for about 25\% of trips performed by driver groups. Other high-demand areas include the North and West Sides, accounts for most amount for several groups. However, Far Southeast Side and Far Southwest Side show significantly lower trip proportions, below 5\% for all driver groups.}
  \label{figure:dropoff_pickup_2019}
\end{figure}
Figures~\ref{figure:dropoff_pickup_2019} and~\ref{figure:dropoff_pickup_2023} indicate that many drivers in both years concentrate on the Central and West Side region. Notably, some drivers show increased airport-focused trips (Driver 8 in 2019 and Driver 9 in 2023), which correlate with \textbf{higher per-trip income} (see Table~\ref{table:earning_driver}). However, driver group which serves the South Side, Far Southwest Side, and Far Southeast Side (Driver Group 8 in 2023) exhibit \textbf{substantially lower average earnings}. Although ride volume in these lower-paying areas has grown from 2019 to 2023, the income gains remain minimal, with driver group 8 reports earnings as low as \$12.74/hour. Our trip assignment simulation algorithm also reveals that, by 2023, a distinct driver group emerged, concentrating on the South Side—a shift not observed in 2019. These disparities highlight how different neighborhoods offer vastly uneven fare opportunities, driven by factors such as longer wait times, perceived safety concerns, and fewer premium trips. Such geographic variations, whether shaped by driver preferences or platform-side decisions, play a crucial role in perpetuating income gaps among drivers.

\subsection{Summary}

Our findings reveal that temporal and regional disparities significantly influence ride-sharing driver earnings. While potential pricing model shifts led to temporary wage increases in early 2021, earnings have since stagnated, resulting in a decline in real, inflation-adjusted income. Moreover, regional earning gaps have widened, with drivers in lower-demand areas earning considerably less than their counterparts in high-traffic regions like the Central and Airport areas.

By clustering our constructed hypothetical drivers, we identify distinct driver groups with unique work patterns, uncovering new spatial trends and emerging low-income driver clusters in 2023. Our results suggest that platform pricing strategies, driver relocation patterns, and algorithmic matching mechanisms are key factors shaping driver income disparities.

\section{Discussions}

\subsection{Transparency in Ride-Sharing Platform Algorithms}
The publicly available Chicago Transportation Network Provider dataset helped us answer many research questions, but ride-sharing platforms still make many of their mechanisms opaque. The lack of transparency in key platform mechanisms---such as pricing models, driver--rider matching algorithms, and driver ranking systems---makes it difficult to pinpoint the exact causes of these disparities. Without greater visibility into these proprietary algorithms, drivers also remain at an information disadvantage, unable to anticipate fare fluctuations or optimize their work schedules effectively.

Pricing models remain opaque, with our analysis revealing that fare adjustments over time have failed to keep pace with inflation, effectively reducing real driver earnings (\cref{sec:results-pricing-stablization}). While platforms advertise dynamic pricing mechanisms that respond to demand surges, drivers have limited insight into how much of the fare they actually receive after platform fees~\cite{santos2020dynamic}. Previous research has shown that drivers tend to work more during peaks for higher compensation~\cite{chen2016dynamic}. A real-time, large-scale understanding of the surge pricing model can help drivers become more informed in planning and organizing their workday, beyond anecdotal observations. Furthermore, researchers can provide prediction models of price surges, helping both drivers and riders adjust plans accordingly. Another key limitation of using the Chicago dataset is the lack of driver earning information. As a result, our analysis can only use the trip fare as a proxy for driver earning. Making such information available can significantly increase transparency into platform operations.

Similarly, the driver-rider matching algorithm remains a black box. Our inferred driver profiles suggest that trip assignments may systematically disadvantage certain groups, particularly those operating in lower-income areas. If the matching algorithm disproportionately favors drivers in high-demand or high-fare regions, it could reinforce existing geographic disparities in earnings. However, such analysis is hard to conduct without access to driver-level information. As discussed in \cref{sec:methods-driver-simulation}, releasing such data may lead to privacy concerns. Our approach is an effort to approximate driver working conditions without needing detailed driver data. However, researchers should still work with ride-sharing platforms to come up with privacy-preserving ways to analyze such data for insights. Also, driver ranking algorithms---which determine access to high-value trips---are equally opaque. While platforms often cite factors such as acceptance rate, customer ratings, and trip history, the lack of public accountability raises concerns regarding potential biases. Accessing such information can support researchers in identifying potential biases, also help drivers provide more desired services to riders.

In all, we call for increased regulatory oversight and platform-level efforts to improve algorithmic transparency. Without clear disclosures on how these systems operate, ride-sharing drivers remain vulnerable to unfair decision-making and fluctuating incomes that they cannot predict or control.

\subsection{Data Analysis Methodology Improvements}
Our study demonstrates the feasibility of simulating reasonable driver profiles from trip-level data, even in the absence of driver-related information. By leveraging a simulation-based approach, we were able to approximate driver earnings, work patterns, and geographic activity. However, there are still areas for improvement for our methodology.

First, a robust evaluation benchmark is needed to validate the accuracy of inferred driver profiles. While our approach provides valuable insights and matches previous empirical findings, the lack of direct ground truth data means we rely on approximations. We need alternative data sources to cross-verify our inferred driver activities. Tools for driver task management, such as Driver's Seat~\cite{calacci2023access}, asks drivers to upload their work tasks and can serve as a potential data source. More autonomous approaches that uses UI understanding techniques and directly collects data from drivers' phones can also scale up this effort~\cite{lu2024crepe}. 

Moreover, expanding the scope of inferred information would provide deeper insights into platform operations. Currently, we infer earnings and work patterns for drivers. Newer algorithms can be developed to analyze additional opaque platform mechanisms as discussed above. Future studies could aim to reconstruct other aspects of opaque platform algorithms, as discussed above, directly from publicly available, large-scale datasets.

Given a large-scale dataset that misses key information aspects, a potential future approach is to self-collect a smaller dataset that contains the necessary details and conduct a joint analysis of both datasets. For example, a smaller dataset that we collect directly from drivers, containing both driver and trip information, can serve both as a benchmark and a basis for use to train machine learning models that predict driver profiles from existing large-scale datasets. Future research can investigate effective measures to combine these different data sources~\cite{harris2018federal} for joint analysis. These methodological advancements can help us to use large-scale ridesharing datasets more effectively and accurately while maintaining driver and rider privacy.

\subsection{Societal Implications: Ride-Sharing as a Reflection of Broader Inequalities}

Our findings revealed regional ride-sharing disparities in the city of Chicago, which largely reflect the broader existing societal inequalities. Drivers working in lower-income neighborhoods---in our case, drivers that service the southern regions of Chicago---consistently earn less, even despite longer work hours. Structural disadvantages, such as lower infrastructure quality, longer wait times, and increased safety concerns---compound the challenges faced by gig workers. Chicago South Side, as a community suffering from violence and poverty, has been an example of social segregation and studied by numerous researchers~\cite{moore2016south, bachin2004building, bell1993community}. As an aspect of a deep-rooted societal issue, ride-sharing inequality in lower-income neighborhoods calls for holistic policymaking efforts from multiple stakeholders.

Our findings provide practical implications for labor activists and policy makers. By providing a more transparent view of drivers’ potential workday experiences, policymakers can better evaluate the labor conditions these platforms create, ensuring that emerging mobility systems align with equity goals. Urban planners and regulators can use these insights to inform policy interventions---such as driver support programs, driver caps, or incentive structures---that promote fairness and mitigate algorithmic biases. Similarly, platform operators themselves might harness these findings to improve their matching algorithms, advancing a more equitable ecosystem that benefits both drivers and passengers.

Research has shown that transportation access can have a positive impact on regional economic growth and productivity~\cite{targa2005economic, banerjee2020road, alstadt2012relationship}. Ride-sharing, as an increasingly critical way of transportation, especially where public transportation is scarce, can support individual and community access to growth opportunities. The persistence of regional earning gaps raises important questions about equity in urban transportation. If ride-sharing platforms are designed primarily to maximize efficiency and revenue, they may inadvertently exacerbate existing economic inequalities by steering high-value rides away from underserved areas~\cite{durand2022access, bocarejo2012transport}.

To address these issues, we call for policy interventions aimed at ensuring fair compensation and equitable access to earning opportunities. Regulators should consider implementing transparency mandates, income stability measures, and algorithmic accountability frameworks to prevent platforms from disproportionately disadvantaging certain driver groups. Moreover, these efforts should be in orchestration with existing efforts to promote infrastructural improvements and public safety in underserved regions. Collaborative initiatives between policymakers, ride-sharing companies, and community organizations can help create a more inclusive transportation ecosystem that benefits both drivers and passengers alike~\cite{baber2022new}.
\section{Limitations and Future Work}

While our study provides valuable insights using both the original dataset and our simulation methods applied to the large-scale Chicago Public Dataset, it is not without limitations. These can be grouped into two categories: (1) limitations related to the analysis of the original dataset, and (2) limitations of the simulation methodology.

First, regarding the original dataset, the lack of sufficient features and detailed information required us to make certain assumptions. For instance, we estimated downtime between trips and assumed that drivers' earnings equaled the trip costs paid by passengers. However, this does not fully capture drivers' actual earnings, as it removes platform fees. To improve the accuracy of earnings estimation, future data collection efforts should include more granular financial details, such as platform fees, driver expenses, and net income.

Second, while our simulation model effectively assign trips to hypothetical drivers and return meaningful insights, it lacks a robust evaluation framework due to the absence of ground-truth data. To validate the findings, we plan to conduct a large-scale qualitative study of driver groups in Chicago. Additionally, due to computational constraints, we limited our analysis to weekly patterns in 2019 and 2023. Future work will focus on enhancing the efficiency of our simulation framework to expand its application to other timeframes, improving its generalizability. Finally, we encourage the development of benchmarks and more quantitative studies to better assess simulation approaches for this type of problem.


\section{Conclusions}
In this study, we explored disparities in ride-sharing earnings and trip distributions using the publicly available Chicago Rideshare dataset (2018–2023). By analyzing both direct observational and a simulation-based methodology, we revealed systematic inequities in driver earnings based on temporal, regional, and algorithmic factors. Our findings reveal that pricing adjustments in recent years have failed to account for inflation, leading to a decline in drivers’ real earnings despite apparent fare stabilization. Additionally, spatial analysis indicates that income gaps have widened over time, with lower-earning zones emerging in Chicago’s South Side and outlying areas.
To address limitations in existing anonymized ride-share datasets, we introduced a simulation-based driver profiling method that reconstructs potential work and earning patterns. This approach allowed us to model driver behaviors, including variations in working hours, trip frequencies, and geographical preferences, which contribute to substantial earnings disparities. Our clustering analysis further revealed the emergence of new driver groups in 2023, suggesting shifts in ride-sharing platform dynamics and potential algorithmic biases in trip allocations.

\bibliographystyle{ACM-Reference-Format}
\bibliography{reference}


\begin{thebibliography}{72}


\ifx \showCODEN    \undefined \def \showCODEN     #1{\unskip}     \fi
\ifx \showDOI      \undefined \def \showDOI       #1{#1}\fi
\ifx \showISBNx    \undefined \def \showISBNx     #1{\unskip}     \fi
\ifx \showISBNxiii \undefined \def \showISBNxiii  #1{\unskip}     \fi
\ifx \showISSN     \undefined \def \showISSN      #1{\unskip}     \fi
\ifx \showLCCN     \undefined \def \showLCCN      #1{\unskip}     \fi
\ifx \shownote     \undefined \def \shownote      #1{#1}          \fi
\ifx \showarticletitle \undefined \def \showarticletitle #1{#1}   \fi
\ifx \showURL      \undefined \def \showURL       {\relax}        \fi
\providecommand\bibfield[2]{#2}
\providecommand\bibinfo[2]{#2}
\providecommand\natexlab[1]{#1}
\providecommand\showeprint[2][]{arXiv:#2}

\bibitem[bk_(2024)]%
        {bk_rideshare_2024}
 \bibinfo{year}{2024}\natexlab{}.
\newblock \showarticletitle{Rideshare Drivers Demand Fair Pay, Treatment in NYC Protest}.
\newblock \bibinfo{journal}{\emph{Brooklyn Reader}} (\bibinfo{year}{2024}).
\newblock
\urldef\tempurl%
\url{https://www.bkreader.com/business-innovation/rideshare-drivers-demand-fair-pay-treatment-in-nyc-protest-9944085}
\showURL{%
\tempurl}
\newblock
\shownote{Accessed: January 26, 2025}.


\bibitem[tru(2024)]%
        {truthout_rideshare_2024}
 \bibinfo{year}{2024}\natexlab{}.
\newblock \showarticletitle{Unionized Rideshare Drivers Vow Future Strikes After Jamming Nashville Airport}.
\newblock \bibinfo{journal}{\emph{Truthout}} (\bibinfo{year}{2024}).
\newblock
\urldef\tempurl%
\url{https://truthout.org/articles/unionized-rideshare-drivers-vow-future-strikes-after-jamming-nashville-airport/}
\showURL{%
\tempurl}
\newblock
\shownote{Accessed: January 26, 2025}.


\bibitem[Allon et~al\mbox{.}(2023)]%
        {allon2023impact}
\bibfield{author}{\bibinfo{person}{Gad Allon}, \bibinfo{person}{Maxime~C Cohen}, {and} \bibinfo{person}{Wichinpong~Park Sinchaisri}.} \bibinfo{year}{2023}\natexlab{}.
\newblock \showarticletitle{The impact of behavioral and economic drivers on gig economy workers}.
\newblock \bibinfo{journal}{\emph{Manufacturing \& Service Operations Management}} \bibinfo{volume}{25}, \bibinfo{number}{4} (\bibinfo{year}{2023}), \bibinfo{pages}{1376--1393}.
\newblock


\bibitem[Alstadt et~al\mbox{.}(2012)]%
        {alstadt2012relationship}
\bibfield{author}{\bibinfo{person}{Brian Alstadt}, \bibinfo{person}{Glen Weisbrod}, {and} \bibinfo{person}{Derek Cutler}.} \bibinfo{year}{2012}\natexlab{}.
\newblock \showarticletitle{Relationship of transportation access and connectivity to local economic outcomes: Statistical analysis}.
\newblock \bibinfo{journal}{\emph{Transportation Research Record}} \bibinfo{volume}{2297}, \bibinfo{number}{1} (\bibinfo{year}{2012}), \bibinfo{pages}{154--162}.
\newblock


\bibitem[Angrist et~al\mbox{.}(2021)]%
        {angrist2021uber}
\bibfield{author}{\bibinfo{person}{Joshua~D Angrist}, \bibinfo{person}{Sydnee Caldwell}, {and} \bibinfo{person}{Jonathan~V Hall}.} \bibinfo{year}{2021}\natexlab{}.
\newblock \showarticletitle{Uber versus taxi: A driver’s eye view}.
\newblock \bibinfo{journal}{\emph{American Economic Journal: Applied Economics}} \bibinfo{volume}{13}, \bibinfo{number}{3} (\bibinfo{year}{2021}), \bibinfo{pages}{272--308}.
\newblock


\bibitem[Baber(2022)]%
        {baber2022new}
\bibfield{author}{\bibinfo{person}{Ashley Baber}.} \bibinfo{year}{2022}\natexlab{}.
\newblock \emph{\bibinfo{title}{The new temporary labor: Regulating the gig economy in Austin, Chicago and New York}}.
\newblock \bibinfo{thesistype}{Ph.\,D. Dissertation}. \bibinfo{school}{Loyola University Chicago}.
\newblock


\bibitem[Bachin(2004)]%
        {bachin2004building}
\bibfield{author}{\bibinfo{person}{Robin~F Bachin}.} \bibinfo{year}{2004}\natexlab{}.
\newblock \bibinfo{booktitle}{\emph{Building the South Side: Urban space and civic culture in Chicago, 1890-1919}}.
\newblock \bibinfo{publisher}{University of Chicago Press}.
\newblock


\bibitem[Bajwa et~al\mbox{.}(2018)]%
        {bajwa2018health}
\bibfield{author}{\bibinfo{person}{Uttam Bajwa}, \bibinfo{person}{Denise Gastaldo}, \bibinfo{person}{Erica Di~Ruggiero}, {and} \bibinfo{person}{Lilian Knorr}.} \bibinfo{year}{2018}\natexlab{}.
\newblock \showarticletitle{The health of workers in the global gig economy}.
\newblock \bibinfo{journal}{\emph{Globalization and health}}  \bibinfo{volume}{14} (\bibinfo{year}{2018}), \bibinfo{pages}{1--4}.
\newblock


\bibitem[Banerjee et~al\mbox{.}(2020)]%
        {banerjee2020road}
\bibfield{author}{\bibinfo{person}{Abhijit Banerjee}, \bibinfo{person}{Esther Duflo}, {and} \bibinfo{person}{Nancy Qian}.} \bibinfo{year}{2020}\natexlab{}.
\newblock \showarticletitle{On the road: Access to transportation infrastructure and economic growth in China}.
\newblock \bibinfo{journal}{\emph{Journal of Development Economics}}  \bibinfo{volume}{145} (\bibinfo{year}{2020}), \bibinfo{pages}{102442}.
\newblock


\bibitem[Banerjee et~al\mbox{.}(2015)]%
        {banerjee2015pricing}
\bibfield{author}{\bibinfo{person}{Siddhartha Banerjee}, \bibinfo{person}{Carlos Riquelme}, {and} \bibinfo{person}{Ramesh Johari}.} \bibinfo{year}{2015}\natexlab{}.
\newblock \showarticletitle{Pricing in ride-share platforms: A queueing-theoretic approach}.
\newblock \bibinfo{journal}{\emph{Available at SSRN 2568258}} (\bibinfo{year}{2015}).
\newblock


\bibitem[Barocas and Selbst(2016)]%
        {barocas2016big}
\bibfield{author}{\bibinfo{person}{Solon Barocas} {and} \bibinfo{person}{Andrew~D Selbst}.} \bibinfo{year}{2016}\natexlab{}.
\newblock \showarticletitle{Big data's disparate impact}.
\newblock \bibinfo{journal}{\emph{Calif. L. Rev.}}  \bibinfo{volume}{104} (\bibinfo{year}{2016}), \bibinfo{pages}{671}.
\newblock


\bibitem[Barrios et~al\mbox{.}(2022)]%
        {barrios2022launching}
\bibfield{author}{\bibinfo{person}{John~M Barrios}, \bibinfo{person}{Yael~V Hochberg}, {and} \bibinfo{person}{Hanyi Yi}.} \bibinfo{year}{2022}\natexlab{}.
\newblock \showarticletitle{Launching with a parachute: The gig economy and new business formation}.
\newblock \bibinfo{journal}{\emph{Journal of Financial Economics}} \bibinfo{volume}{144}, \bibinfo{number}{1} (\bibinfo{year}{2022}), \bibinfo{pages}{22--43}.
\newblock


\bibitem[Bell and Jenkins(1993)]%
        {bell1993community}
\bibfield{author}{\bibinfo{person}{Carl~C Bell} {and} \bibinfo{person}{Esther~J Jenkins}.} \bibinfo{year}{1993}\natexlab{}.
\newblock \showarticletitle{Community violence and children on Chicago’s southside}.
\newblock \bibinfo{journal}{\emph{Psychiatry}} \bibinfo{volume}{56}, \bibinfo{number}{1} (\bibinfo{year}{1993}), \bibinfo{pages}{46--54}.
\newblock


\bibitem[Berg(2015)]%
        {berg2015income}
\bibfield{author}{\bibinfo{person}{Janine Berg}.} \bibinfo{year}{2015}\natexlab{}.
\newblock \showarticletitle{Income security in the on-demand economy: Findings and policy lessons from a survey of crowdworkers}.
\newblock \bibinfo{journal}{\emph{Comp. Lab. L. \& Pol'y J.}}  \bibinfo{volume}{37} (\bibinfo{year}{2015}), \bibinfo{pages}{543}.
\newblock


\bibitem[Berger et~al\mbox{.}(2019)]%
        {berger2019uber}
\bibfield{author}{\bibinfo{person}{Thor Berger}, \bibinfo{person}{Carl~Benedikt Frey}, \bibinfo{person}{Guy Levin}, {and} \bibinfo{person}{Santosh~Rao Danda}.} \bibinfo{year}{2019}\natexlab{}.
\newblock \showarticletitle{Uber happy? Work and well-being in the ‘gig economy’}.
\newblock \bibinfo{journal}{\emph{Economic Policy}} \bibinfo{volume}{34}, \bibinfo{number}{99} (\bibinfo{year}{2019}), \bibinfo{pages}{429--477}.
\newblock


\bibitem[Bocarejo~S and Oviedo~H(2012)]%
        {bocarejo2012transport}
\bibfield{author}{\bibinfo{person}{Juan~Pablo Bocarejo~S} {and} \bibinfo{person}{Daniel~Ricardo Oviedo~H}.} \bibinfo{year}{2012}\natexlab{}.
\newblock \showarticletitle{Transport accessibility and social inequities: a tool for identification of mobility needs and evaluation of transport investments}.
\newblock \bibinfo{journal}{\emph{Journal of transport geography}}  \bibinfo{volume}{24} (\bibinfo{year}{2012}), \bibinfo{pages}{142--154}.
\newblock


\bibitem[Brown(2024)]%
        {brown2024driving}
\bibfield{author}{\bibinfo{person}{Anne Brown}.} \bibinfo{year}{2024}\natexlab{}.
\newblock \bibinfo{booktitle}{\emph{Driving to Opportunity? Work and Car Access Among Low-Income Ride-Hail and Delivery Drivers}}.
\newblock \bibinfo{type}{{T}echnical {R}eport}. \bibinfo{institution}{Center for Open Science}.
\newblock


\bibitem[Calacci and Stein(2023)]%
        {calacci2023access}
\bibfield{author}{\bibinfo{person}{Dan Calacci} {and} \bibinfo{person}{Jake Stein}.} \bibinfo{year}{2023}\natexlab{}.
\newblock \showarticletitle{From access to understanding: Collective data governance for workers}.
\newblock \bibinfo{journal}{\emph{European Labour Law Journal}} \bibinfo{volume}{14}, \bibinfo{number}{2} (\bibinfo{year}{2023}), \bibinfo{pages}{253--282}.
\newblock


\bibitem[{California Secretary of State}(2020)]%
        {california_prop22_2020}
\bibfield{author}{\bibinfo{person}{{California Secretary of State}}.} \bibinfo{year}{2020}\natexlab{}.
\newblock \bibinfo{title}{California Proposition 22, App-Based Drivers as Contractors and Labor Policies Initiative}.
\newblock \bibinfo{howpublished}{\url{https://vig.cdn.sos.ca.gov/2020/general/pdf/topl-prop22.pdf}}.
\newblock
\newblock
\shownote{Accessed: January 26, 2025}.


\bibitem[Cao et~al\mbox{.}(2021)]%
        {cao2021optimization}
\bibfield{author}{\bibinfo{person}{Yi Cao}, \bibinfo{person}{Shan Wang}, {and} \bibinfo{person}{Jinyang Li}.} \bibinfo{year}{2021}\natexlab{}.
\newblock \showarticletitle{The optimization model of ride-sharing route for ride hailing considering both system optimization and user fairness}.
\newblock \bibinfo{journal}{\emph{Sustainability}} \bibinfo{volume}{13}, \bibinfo{number}{2} (\bibinfo{year}{2021}), \bibinfo{pages}{902}.
\newblock


\bibitem[Chan and Shaheen(2012)]%
        {chan2012ridesharing}
\bibfield{author}{\bibinfo{person}{Nelson~D Chan} {and} \bibinfo{person}{Susan~A Shaheen}.} \bibinfo{year}{2012}\natexlab{}.
\newblock \showarticletitle{Ridesharing in North America: Past, present, and future}.
\newblock \bibinfo{journal}{\emph{Transport reviews}} \bibinfo{volume}{32}, \bibinfo{number}{1} (\bibinfo{year}{2012}), \bibinfo{pages}{93--112}.
\newblock


\bibitem[Chen et~al\mbox{.}(2019)]%
        {chen2019value}
\bibfield{author}{\bibinfo{person}{M~Keith Chen}, \bibinfo{person}{Peter~E Rossi}, \bibinfo{person}{Judith~A Chevalier}, {and} \bibinfo{person}{Emily Oehlsen}.} \bibinfo{year}{2019}\natexlab{}.
\newblock \showarticletitle{The value of flexible work: Evidence from Uber drivers}.
\newblock \bibinfo{journal}{\emph{Journal of political economy}} \bibinfo{volume}{127}, \bibinfo{number}{6} (\bibinfo{year}{2019}), \bibinfo{pages}{2735--2794}.
\newblock


\bibitem[Chen and Sheldon(2016)]%
        {chen2016dynamic}
\bibfield{author}{\bibinfo{person}{M~Keith Chen} {and} \bibinfo{person}{Michael Sheldon}.} \bibinfo{year}{2016}\natexlab{}.
\newblock \showarticletitle{Dynamic pricing in a labor market: Surge pricing and flexible work on the Uber platform.}
\newblock \bibinfo{journal}{\emph{Ec}}  \bibinfo{volume}{16} (\bibinfo{year}{2016}), \bibinfo{pages}{455}.
\newblock


\bibitem[{City of Chicago}(2023a)]%
        {chicago_tnp_2018}
\bibfield{author}{\bibinfo{person}{{City of Chicago}}.} \bibinfo{year}{2023}\natexlab{a}.
\newblock \bibinfo{title}{Transportation Network Providers - Trips 2018-2022}.
\newblock \bibinfo{howpublished}{\url{https://data.cityofchicago.org/Transportation/Transportation-Network-Providers-Trips-2018-2022-/m6dm-c72p/}}.
\newblock


\bibitem[{City of Chicago}(2023b)]%
        {chicago_tnp_2023}
\bibfield{author}{\bibinfo{person}{{City of Chicago}}.} \bibinfo{year}{2023}\natexlab{b}.
\newblock \bibinfo{title}{Transportation Network Providers - Trips 2023}.
\newblock \bibinfo{howpublished}{\url{https://data.cityofchicago.org/Transportation/Transportation-Network-Providers-Trips-2023-/n26f-ihde}}.
\newblock
\newblock
\shownote{Accessed: January 26, 2025}.


\bibitem[Cook et~al\mbox{.}(2021)]%
        {cook2021gender}
\bibfield{author}{\bibinfo{person}{Cody Cook}, \bibinfo{person}{Rebecca Diamond}, \bibinfo{person}{Jonathan~V Hall}, \bibinfo{person}{John~A List}, {and} \bibinfo{person}{Paul Oyer}.} \bibinfo{year}{2021}\natexlab{}.
\newblock \showarticletitle{The gender earnings gap in the gig economy: Evidence from over a million rideshare drivers}.
\newblock \bibinfo{journal}{\emph{The Review of Economic Studies}} \bibinfo{volume}{88}, \bibinfo{number}{5} (\bibinfo{year}{2021}), \bibinfo{pages}{2210--2238}.
\newblock


\bibitem[Cram et~al\mbox{.}(2022)]%
        {cram2022examining}
\bibfield{author}{\bibinfo{person}{W~Alec Cram}, \bibinfo{person}{Martin Wiener}, \bibinfo{person}{Monideepa Tarafdar}, {and} \bibinfo{person}{Alexander Benlian}.} \bibinfo{year}{2022}\natexlab{}.
\newblock \showarticletitle{Examining the impact of algorithmic control on Uber drivers’ technostress}.
\newblock \bibinfo{journal}{\emph{Journal of management information systems}} \bibinfo{volume}{39}, \bibinfo{number}{2} (\bibinfo{year}{2022}), \bibinfo{pages}{426--453}.
\newblock


\bibitem[Cramer and Krueger(2016a)]%
        {NBERw22083}
\bibfield{author}{\bibinfo{person}{Judd Cramer} {and} \bibinfo{person}{Alan~B Krueger}.} \bibinfo{year}{2016}\natexlab{a}.
\newblock \bibinfo{booktitle}{\emph{Disruptive Change in the Taxi Business: The Case of Uber}}.
\newblock \bibinfo{type}{Working Paper} 22083. \bibinfo{institution}{National Bureau of Economic Research}.
\newblock
\urldef\tempurl%
\url{https://doi.org/10.3386/w22083}
\showDOI{\tempurl}


\bibitem[Cramer and Krueger(2016b)]%
        {cramer2016disruptive}
\bibfield{author}{\bibinfo{person}{Judd Cramer} {and} \bibinfo{person}{Alan~B Krueger}.} \bibinfo{year}{2016}\natexlab{b}.
\newblock \showarticletitle{Disruptive change in the taxi business: The case of Uber}.
\newblock \bibinfo{journal}{\emph{American Economic Review}} \bibinfo{volume}{106}, \bibinfo{number}{5} (\bibinfo{year}{2016}), \bibinfo{pages}{177--182}.
\newblock


\bibitem[de~Ruijter et~al\mbox{.}(2024)]%
        {de2024ridesourcing}
\bibfield{author}{\bibinfo{person}{Arjan de Ruijter}, \bibinfo{person}{Oded Cats}, {and} \bibinfo{person}{Hans van Lint}.} \bibinfo{year}{2024}\natexlab{}.
\newblock \showarticletitle{Ridesourcing platforms thrive on socio-economic inequality}.
\newblock \bibinfo{journal}{\emph{Scientific Reports}} \bibinfo{volume}{14}, \bibinfo{number}{1} (\bibinfo{year}{2024}), \bibinfo{pages}{7371}.
\newblock


\bibitem[De~Stefano(2015)]%
        {de2015rise}
\bibfield{author}{\bibinfo{person}{Valerio De~Stefano}.} \bibinfo{year}{2015}\natexlab{}.
\newblock \showarticletitle{The rise of the just-in-time workforce: On-demand work, crowdwork, and labor protection in the gig-economy}.
\newblock \bibinfo{journal}{\emph{Comp. Lab. L. \& Pol'y J.}}  \bibinfo{volume}{37} (\bibinfo{year}{2015}), \bibinfo{pages}{471}.
\newblock


\bibitem[Di~Febbraro et~al\mbox{.}(2013)]%
        {di2013optimization}
\bibfield{author}{\bibinfo{person}{Angela Di~Febbraro}, \bibinfo{person}{Enrico Gattorna}, {and} \bibinfo{person}{Nicola Sacco}.} \bibinfo{year}{2013}\natexlab{}.
\newblock \showarticletitle{Optimization of dynamic ridesharing systems}.
\newblock \bibinfo{journal}{\emph{Transportation research record}} \bibinfo{volume}{2359}, \bibinfo{number}{1} (\bibinfo{year}{2013}), \bibinfo{pages}{44--50}.
\newblock


\bibitem[Duggan et~al\mbox{.}(2023)]%
        {duggan2023algorithmic}
\bibfield{author}{\bibinfo{person}{James Duggan}, \bibinfo{person}{Ronan Carbery}, \bibinfo{person}{Anthony McDonnell}, {and} \bibinfo{person}{Ultan Sherman}.} \bibinfo{year}{2023}\natexlab{}.
\newblock \showarticletitle{Algorithmic HRM control in the gig economy: The app-worker perspective}.
\newblock \bibinfo{journal}{\emph{Human Resource Management}} \bibinfo{volume}{62}, \bibinfo{number}{6} (\bibinfo{year}{2023}), \bibinfo{pages}{883--899}.
\newblock


\bibitem[Duggan et~al\mbox{.}(2020)]%
        {duggan2020algorithmic}
\bibfield{author}{\bibinfo{person}{James Duggan}, \bibinfo{person}{Ultan Sherman}, \bibinfo{person}{Ronan Carbery}, {and} \bibinfo{person}{Anthony McDonnell}.} \bibinfo{year}{2020}\natexlab{}.
\newblock \showarticletitle{Algorithmic management and app-work in the gig economy: A research agenda for employment relations and HRM}.
\newblock \bibinfo{journal}{\emph{Human Resource Management Journal}} \bibinfo{volume}{30}, \bibinfo{number}{1} (\bibinfo{year}{2020}), \bibinfo{pages}{114--132}.
\newblock


\bibitem[Durand et~al\mbox{.}(2022)]%
        {durand2022access}
\bibfield{author}{\bibinfo{person}{Anne Durand}, \bibinfo{person}{Toon Zijlstra}, \bibinfo{person}{Niels van Oort}, \bibinfo{person}{Sascha Hoogendoorn-Lanser}, {and} \bibinfo{person}{Serge Hoogendoorn}.} \bibinfo{year}{2022}\natexlab{}.
\newblock \showarticletitle{Access denied? Digital inequality in transport services}.
\newblock \bibinfo{journal}{\emph{Transport Reviews}} \bibinfo{volume}{42}, \bibinfo{number}{1} (\bibinfo{year}{2022}), \bibinfo{pages}{32--57}.
\newblock


\bibitem[Ge et~al\mbox{.}(2020)]%
        {GE2020104205}
\bibfield{author}{\bibinfo{person}{Yanbo Ge}, \bibinfo{person}{Christopher~R. Knittel}, \bibinfo{person}{Don MacKenzie}, {and} \bibinfo{person}{Stephen Zoepf}.} \bibinfo{year}{2020}\natexlab{}.
\newblock \showarticletitle{Racial discrimination in transportation network companies}.
\newblock \bibinfo{journal}{\emph{Journal of Public Economics}}  \bibinfo{volume}{190} (\bibinfo{year}{2020}), \bibinfo{pages}{104205}.
\newblock
\showISSN{0047-2727}
\urldef\tempurl%
\url{https://doi.org/10.1016/j.jpubeco.2020.104205}
\showDOI{\tempurl}


\bibitem[Hall and Krueger(2018)]%
        {hall2018analysis}
\bibfield{author}{\bibinfo{person}{Jonathan~V Hall} {and} \bibinfo{person}{Alan~B Krueger}.} \bibinfo{year}{2018}\natexlab{}.
\newblock \showarticletitle{An analysis of the labor market for Uber’s driver-partners in the United States}.
\newblock \bibinfo{journal}{\emph{Ilr Review}} \bibinfo{volume}{71}, \bibinfo{number}{3} (\bibinfo{year}{2018}), \bibinfo{pages}{705--732}.
\newblock


\bibitem[Harris-Kojetin and Groves(2018)]%
        {harris2018federal}
\bibfield{author}{\bibinfo{person}{Brian~A Harris-Kojetin} {and} \bibinfo{person}{Robert~M Groves}.} \bibinfo{year}{2018}\natexlab{}.
\newblock \showarticletitle{Federal statistics, multiple data sources, and privacy protection: next steps}.
\newblock  (\bibinfo{year}{2018}).
\newblock


\bibitem[Henao and Marshall(2019)]%
        {RePEc:kap:transp:v:46:y:2019:i:6:d:10.1007_s11116-018-9923-2}
\bibfield{author}{\bibinfo{person}{Alejandro Henao} {and} \bibinfo{person}{Wesley~E. Marshall}.} \bibinfo{year}{2019}\natexlab{}.
\newblock \showarticletitle{{The impact of ride-hailing on vehicle miles traveled}}.
\newblock \bibinfo{journal}{\emph{Transportation}} \bibinfo{volume}{46}, \bibinfo{number}{6} (\bibinfo{date}{December} \bibinfo{year}{2019}), \bibinfo{pages}{2173--2194}.
\newblock
\urldef\tempurl%
\url{https://doi.org/10.1007/s11116-018-9923-2}
\showDOI{\tempurl}


\bibitem[Kadolkar et~al\mbox{.}(2024)]%
        {kadolkar2024algorithmic}
\bibfield{author}{\bibinfo{person}{Imran Kadolkar}, \bibinfo{person}{Sven Kepes}, {and} \bibinfo{person}{Mahesh Subramony}.} \bibinfo{year}{2024}\natexlab{}.
\newblock \showarticletitle{Algorithmic management in the gig economy: A systematic review and research integration}.
\newblock \bibinfo{journal}{\emph{Journal of Organizational Behavior}} (\bibinfo{year}{2024}).
\newblock


\bibitem[Kloostra(2022)]%
        {kloostra2022algorithmic}
\bibfield{author}{\bibinfo{person}{Jorn Kloostra}.} \bibinfo{year}{2022}\natexlab{}.
\newblock \showarticletitle{Algorithmic pricing: A concern for platform workers?}
\newblock \bibinfo{journal}{\emph{European Labour Law Journal}} \bibinfo{volume}{13}, \bibinfo{number}{1} (\bibinfo{year}{2022}), \bibinfo{pages}{108--126}.
\newblock


\bibitem[Kucharski and Cats(2022)]%
        {kucharski2022simulating}
\bibfield{author}{\bibinfo{person}{Rafa{\l} Kucharski} {and} \bibinfo{person}{Oded Cats}.} \bibinfo{year}{2022}\natexlab{}.
\newblock \showarticletitle{Simulating two-sided mobility platforms with MaaSSim}.
\newblock \bibinfo{journal}{\emph{Plos one}} \bibinfo{volume}{17}, \bibinfo{number}{6} (\bibinfo{year}{2022}), \bibinfo{pages}{e0269682}.
\newblock


\bibitem[Kumar et~al\mbox{.}(2023)]%
        {kumar2023using}
\bibfield{author}{\bibinfo{person}{Ashwin Kumar}, \bibinfo{person}{Yevgeniy Vorobeychik}, {and} \bibinfo{person}{William Yeoh}.} \bibinfo{year}{2023}\natexlab{}.
\newblock \showarticletitle{Using simple incentives to improve two-sided fairness in ridesharing systems}. In \bibinfo{booktitle}{\emph{Proceedings of the International Conference on Automated Planning and Scheduling}}, Vol.~\bibinfo{volume}{33}. \bibinfo{pages}{227--235}.
\newblock


\bibitem[Lee et~al\mbox{.}(2015)]%
        {lee2015working}
\bibfield{author}{\bibinfo{person}{Min~Kyung Lee}, \bibinfo{person}{Daniel Kusbit}, \bibinfo{person}{Evan Metsky}, {and} \bibinfo{person}{Laura Dabbish}.} \bibinfo{year}{2015}\natexlab{}.
\newblock \showarticletitle{Working with machines: The impact of algorithmic and data-driven management on human workers}. In \bibinfo{booktitle}{\emph{Proceedings of the 33rd annual ACM conference on human factors in computing systems}}. \bibinfo{pages}{1603--1612}.
\newblock


\bibitem[Liu and Li(2023)]%
        {liu2023economic}
\bibfield{author}{\bibinfo{person}{Yang Liu} {and} \bibinfo{person}{Sen Li}.} \bibinfo{year}{2023}\natexlab{}.
\newblock \showarticletitle{An economic analysis of on-demand food delivery platforms: Impacts of regulations and integration with ride-sourcing platforms}.
\newblock \bibinfo{journal}{\emph{Transportation Research Part E: Logistics and Transportation Review}}  \bibinfo{volume}{171} (\bibinfo{year}{2023}), \bibinfo{pages}{103019}.
\newblock


\bibitem[Liu et~al\mbox{.}(2024)]%
        {liu2024evaluating}
\bibfield{author}{\bibinfo{person}{Yuhan Liu}, \bibinfo{person}{Yuhan Zheng}, \bibinfo{person}{Siyuan Zhang}, {and} \bibinfo{person}{Lydia~T Liu}.} \bibinfo{year}{2024}\natexlab{}.
\newblock \showarticletitle{Evaluating Fairness in Black-box Algorithmic Markets: A Case Study of Ride Sharing in Chicago}.
\newblock \bibinfo{journal}{\emph{arXiv preprint arXiv:2407.20522}} (\bibinfo{year}{2024}).
\newblock


\bibitem[Lloyd(1982)]%
        {lloyd1982least}
\bibfield{author}{\bibinfo{person}{Stuart Lloyd}.} \bibinfo{year}{1982}\natexlab{}.
\newblock \showarticletitle{Least squares quantization in PCM}.
\newblock \bibinfo{journal}{\emph{IEEE transactions on information theory}} \bibinfo{volume}{28}, \bibinfo{number}{2} (\bibinfo{year}{1982}), \bibinfo{pages}{129--137}.
\newblock


\bibitem[Lobel(2017)]%
        {lobel2017gig}
\bibfield{author}{\bibinfo{person}{Orly Lobel}.} \bibinfo{year}{2017}\natexlab{}.
\newblock \showarticletitle{The gig economy \& the future of employment and labor law}.
\newblock \bibinfo{journal}{\emph{USFL Rev.}}  \bibinfo{volume}{51} (\bibinfo{year}{2017}), \bibinfo{pages}{51}.
\newblock


\bibitem[Lu et~al\mbox{.}(2024)]%
        {lu2024crepe}
\bibfield{author}{\bibinfo{person}{Yuwen Lu}, \bibinfo{person}{Meng Chen}, \bibinfo{person}{Qi Zhao}, \bibinfo{person}{Victor Cox}, \bibinfo{person}{Yang Yang}, \bibinfo{person}{Meng Jiang}, \bibinfo{person}{Jay Brockman}, \bibinfo{person}{Tamara Kay}, {and} \bibinfo{person}{Toby Jia-Jun Li}.} \bibinfo{year}{2024}\natexlab{}.
\newblock \showarticletitle{Crepe: A Mobile Screen Data Collector Using Graph Query}.
\newblock \bibinfo{journal}{\emph{arXiv preprint arXiv:2406.16173}} (\bibinfo{year}{2024}).
\newblock


\bibitem[Manzo~IV et~al\mbox{.}(2022)]%
        {manzo2022improving}
\bibfield{author}{\bibinfo{person}{Frank Manzo~IV}, \bibinfo{person}{Larissa Petrucci}, {and} \bibinfo{person}{Robert Bruno}.} \bibinfo{year}{2022}\natexlab{}.
\newblock \bibinfo{title}{Improving Labor Standards for Uber and Lyft Drivers in Chicago: Classifying Drivers as Employees Would Deliver Superior Outcomes}.
\newblock
\newblock
\urldef\tempurl%
\url{https://lep.illinois.edu/publications/improving-labor-standards-for-uber-and-lyft-drivers-in-chicago-classifying-drivers-as-employees-would-deliver-superior-outcomes/}
\showURL{%
\tempurl}


\bibitem[Mishel(2018)]%
        {mishel2018uber}
\bibfield{author}{\bibinfo{person}{Lawrence Mishel}.} \bibinfo{year}{2018}\natexlab{}.
\newblock \showarticletitle{Uber and the labor market: Uber drivers’ compensation, wages, and the scale of Uber and the gig economy}.
\newblock  (\bibinfo{year}{2018}).
\newblock


\bibitem[Moore(2016)]%
        {moore2016south}
\bibfield{author}{\bibinfo{person}{Natalie~Y Moore}.} \bibinfo{year}{2016}\natexlab{}.
\newblock \bibinfo{booktitle}{\emph{The south side: A portrait of Chicago and American segregation}}.
\newblock \bibinfo{publisher}{Macmillan}.
\newblock


\bibitem[Nanda et~al\mbox{.}(2020)]%
        {nanda2020balancing}
\bibfield{author}{\bibinfo{person}{Vedant Nanda}, \bibinfo{person}{Pan Xu}, \bibinfo{person}{Karthik~Abhinav Sankararaman}, \bibinfo{person}{John Dickerson}, {and} \bibinfo{person}{Aravind Srinivasan}.} \bibinfo{year}{2020}\natexlab{}.
\newblock \showarticletitle{Balancing the tradeoff between profit and fairness in rideshare platforms during high-demand hours}. In \bibinfo{booktitle}{\emph{Proceedings of the AAAI conference on artificial intelligence}}, Vol.~\bibinfo{volume}{34}. \bibinfo{pages}{2210--2217}.
\newblock


\bibitem[of~Labor~Statistics(2025)]%
        {BLS_CPI}
\bibfield{author}{\bibinfo{person}{Bureau of Labor~Statistics}.} \bibinfo{year}{2025}\natexlab{}.
\newblock \bibinfo{title}{Consumer Price Index for All Urban Consumers: Chicago Area}.
\newblock \bibinfo{howpublished}{\url{https://data.bls.gov/timeseries/CUURS23ASA0?amp\%253bdata_tool=XGtable&output_view=data&include_graphs=true}}.
\newblock
\newblock
\shownote{Accessed: February 7, 2025}.


\bibitem[Rayle et~al\mbox{.}(2016)]%
        {RAYLE2016168}
\bibfield{author}{\bibinfo{person}{Lisa Rayle}, \bibinfo{person}{Danielle Dai}, \bibinfo{person}{Nelson Chan}, \bibinfo{person}{Robert Cervero}, {and} \bibinfo{person}{Susan Shaheen}.} \bibinfo{year}{2016}\natexlab{}.
\newblock \showarticletitle{Just a better taxi? A survey-based comparison of taxis, transit, and ridesourcing services in San Francisco}.
\newblock \bibinfo{journal}{\emph{Transport Policy}}  \bibinfo{volume}{45} (\bibinfo{year}{2016}), \bibinfo{pages}{168--178}.
\newblock
\showISSN{0967-070X}
\urldef\tempurl%
\url{https://doi.org/10.1016/j.tranpol.2015.10.004}
\showDOI{\tempurl}


\bibitem[Rosenblat(2018)]%
        {rosenblat2018uberland}
\bibfield{author}{\bibinfo{person}{Alex Rosenblat}.} \bibinfo{year}{2018}\natexlab{}.
\newblock \bibinfo{booktitle}{\emph{Uberland: How algorithms are rewriting the rules of work}}.
\newblock \bibinfo{publisher}{Univ of California Press}.
\newblock


\bibitem[Rosenblat and Stark(2016)]%
        {rosenblat2016algorithmic}
\bibfield{author}{\bibinfo{person}{Alex Rosenblat} {and} \bibinfo{person}{Luke Stark}.} \bibinfo{year}{2016}\natexlab{}.
\newblock \showarticletitle{Algorithmic labor and information asymmetries: A case study of Uber’s drivers}.
\newblock \bibinfo{journal}{\emph{International journal of communication}}  \bibinfo{volume}{10} (\bibinfo{year}{2016}), \bibinfo{pages}{27}.
\newblock


\bibitem[Rousseeuw(1987)]%
        {rousseeuw1987silhouettes}
\bibfield{author}{\bibinfo{person}{Peter~J Rousseeuw}.} \bibinfo{year}{1987}\natexlab{}.
\newblock \showarticletitle{Silhouettes: a graphical aid to the interpretation and validation of cluster analysis}.
\newblock \bibinfo{journal}{\emph{Journal of computational and applied mathematics}}  \bibinfo{volume}{20} (\bibinfo{year}{1987}), \bibinfo{pages}{53--65}.
\newblock


\bibitem[Rovatsos et~al\mbox{.}(2019)]%
        {rovatsos2019landscape}
\bibfield{author}{\bibinfo{person}{Michael Rovatsos}, \bibinfo{person}{Brent Mittelstadt}, {and} \bibinfo{person}{Ansgar Koene}.} \bibinfo{year}{2019}\natexlab{}.
\newblock \showarticletitle{Landscape summary: Bias in algorithmic decision-making: What is bias in algorithmic decision-making, how can we identify it, and how can we mitigate it?}
\newblock  (\bibinfo{year}{2019}).
\newblock


\bibitem[Ruch et~al\mbox{.}(2020)]%
        {ruch2020quantifying}
\bibfield{author}{\bibinfo{person}{Claudio Ruch}, \bibinfo{person}{ChengQi Lu}, \bibinfo{person}{Lukas Sieber}, {and} \bibinfo{person}{Emilio Frazzoli}.} \bibinfo{year}{2020}\natexlab{}.
\newblock \showarticletitle{Quantifying the efficiency of ride sharing}.
\newblock \bibinfo{journal}{\emph{IEEE Transactions on Intelligent Transportation Systems}} \bibinfo{volume}{22}, \bibinfo{number}{9} (\bibinfo{year}{2020}), \bibinfo{pages}{5811--5816}.
\newblock


\bibitem[Santos et~al\mbox{.}(2020)]%
        {santos2020dynamic}
\bibfield{author}{\bibinfo{person}{Flavio Andrew do~Nascimento Santos}, \bibinfo{person}{Ver{\^o}nica~Feder Mayer}, {and} \bibinfo{person}{Osiris Ricardo~Bezerra Marques}.} \bibinfo{year}{2020}\natexlab{}.
\newblock \showarticletitle{Dynamic pricing and price fairness perceptions: a study of the use of the Uber app in travels}.
\newblock \bibinfo{journal}{\emph{Turismo: Vis{\~a}o e A{\c{c}}{\~a}o}}  \bibinfo{volume}{21} (\bibinfo{year}{2020}), \bibinfo{pages}{239--264}.
\newblock


\bibitem[Schreieck et~al\mbox{.}(2016)]%
        {schreieck2016matching}
\bibfield{author}{\bibinfo{person}{Maximilian Schreieck}, \bibinfo{person}{Hazem Safetli}, \bibinfo{person}{Sajjad~Ali Siddiqui}, \bibinfo{person}{Christoph Pfl{\"u}gler}, \bibinfo{person}{Manuel Wiesche}, {and} \bibinfo{person}{Helmut Krcmar}.} \bibinfo{year}{2016}\natexlab{}.
\newblock \showarticletitle{A matching algorithm for dynamic ridesharing}.
\newblock \bibinfo{journal}{\emph{Transportation Research Procedia}}  \bibinfo{volume}{19} (\bibinfo{year}{2016}), \bibinfo{pages}{272--285}.
\newblock


\bibitem[Shapiro(2020)]%
        {shapiro2020dynamic}
\bibfield{author}{\bibinfo{person}{Aaron Shapiro}.} \bibinfo{year}{2020}\natexlab{}.
\newblock \showarticletitle{Dynamic exploits: calculative asymmetries in the on-demand economy}.
\newblock \bibinfo{journal}{\emph{New Technology, Work and Employment}} \bibinfo{volume}{35}, \bibinfo{number}{2} (\bibinfo{year}{2020}), \bibinfo{pages}{162--177}.
\newblock


\bibitem[Tan et~al\mbox{.}(2021)]%
        {tan2021ethical}
\bibfield{author}{\bibinfo{person}{Zhi~Ming Tan}, \bibinfo{person}{Nikita Aggarwal}, \bibinfo{person}{Josh Cowls}, \bibinfo{person}{Jessica Morley}, \bibinfo{person}{Mariarosaria Taddeo}, {and} \bibinfo{person}{Luciano Floridi}.} \bibinfo{year}{2021}\natexlab{}.
\newblock \showarticletitle{The ethical debate about the gig economy: A review and critical analysis}.
\newblock \bibinfo{journal}{\emph{Technology in Society}}  \bibinfo{volume}{65} (\bibinfo{year}{2021}), \bibinfo{pages}{101594}.
\newblock


\bibitem[Targa et~al\mbox{.}(2005)]%
        {targa2005economic}
\bibfield{author}{\bibinfo{person}{Felipe Targa}, \bibinfo{person}{Kelly~J Clifton}, {and} \bibinfo{person}{Hani~S Mahmassani}.} \bibinfo{year}{2005}\natexlab{}.
\newblock \showarticletitle{Economic activity and transportation access: an econometric analysis of business spatial patterns}.
\newblock \bibinfo{journal}{\emph{Transportation research record}} \bibinfo{volume}{1932}, \bibinfo{number}{1} (\bibinfo{year}{2005}), \bibinfo{pages}{61--71}.
\newblock


\bibitem[{Uber Technologies, Inc.}(2023)]%
        {uber_financials_2023}
\bibfield{author}{\bibinfo{person}{{Uber Technologies, Inc.}}} \bibinfo{year}{2023}\natexlab{}.
\newblock \bibinfo{title}{Financials}.
\newblock \bibinfo{howpublished}{\url{https://investor.uber.com/financials/default.aspx}}.
\newblock
\newblock
\shownote{Accessed: January 26, 2025}.


\bibitem[Van~Doorn et~al\mbox{.}(2020)]%
        {van2020wage}
\bibfield{author}{\bibinfo{person}{Niels Van~Doorn} {et~al\mbox{.}}} \bibinfo{year}{2020}\natexlab{}.
\newblock \showarticletitle{From a wage to a wager: Dynamic pricing in the gig economy}.
\newblock \bibinfo{journal}{\emph{Platform Equality}} (\bibinfo{year}{2020}).
\newblock


\bibitem[Wood et~al\mbox{.}(2019)]%
        {wood2019good}
\bibfield{author}{\bibinfo{person}{Alex~J Wood}, \bibinfo{person}{Mark Graham}, \bibinfo{person}{Vili Lehdonvirta}, {and} \bibinfo{person}{Isis Hjorth}.} \bibinfo{year}{2019}\natexlab{}.
\newblock \showarticletitle{Good gig, bad gig: autonomy and algorithmic control in the global gig economy}.
\newblock \bibinfo{journal}{\emph{Work, employment and society}} \bibinfo{volume}{33}, \bibinfo{number}{1} (\bibinfo{year}{2019}), \bibinfo{pages}{56--75}.
\newblock


\bibitem[Zhang et~al\mbox{.}(2022)]%
        {zhang2022algorithmic}
\bibfield{author}{\bibinfo{person}{Angie Zhang}, \bibinfo{person}{Alexander Boltz}, \bibinfo{person}{Chun~Wei Wang}, {and} \bibinfo{person}{Min~Kyung Lee}.} \bibinfo{year}{2022}\natexlab{}.
\newblock \showarticletitle{Algorithmic management reimagined for workers and by workers: Centering worker well-being in gig work}. In \bibinfo{booktitle}{\emph{Proceedings of the 2022 CHI conference on human factors in computing systems}}. \bibinfo{pages}{1--20}.
\newblock


\bibitem[Zhang et~al\mbox{.}(2024)]%
        {zhang2024data}
\bibfield{author}{\bibinfo{person}{Angie Zhang}, \bibinfo{person}{Rocita Rana}, \bibinfo{person}{Alexander Boltz}, \bibinfo{person}{Veena Dubal}, {and} \bibinfo{person}{Min~Kyung Lee}.} \bibinfo{year}{2024}\natexlab{}.
\newblock \showarticletitle{Data Probes as Boundary Objects for Technology Policy Design: Demystifying Technology for Policymakers and Aligning Stakeholder Objectives in Rideshare Gig Work}. In \bibinfo{booktitle}{\emph{Proceedings of the CHI Conference on Human Factors in Computing Systems}}. \bibinfo{pages}{1--21}.
\newblock


\bibitem[Zhang et~al\mbox{.}(2020)]%
        {zhang2020pricing}
\bibfield{author}{\bibinfo{person}{Chaoli Zhang}, \bibinfo{person}{Jiapeng Xie}, \bibinfo{person}{Fan Wu}, \bibinfo{person}{Xiaofeng Gao}, {and} \bibinfo{person}{Guihai Chen}.} \bibinfo{year}{2020}\natexlab{}.
\newblock \showarticletitle{Pricing and allocation algorithm designs in dynamic ridesharing system}.
\newblock \bibinfo{journal}{\emph{Theoretical Computer Science}}  \bibinfo{volume}{803} (\bibinfo{year}{2020}), \bibinfo{pages}{94--104}.
\newblock


\bibitem[Zhu et~al\mbox{.}(2024)]%
        {zhu2024gig}
\bibfield{author}{\bibinfo{person}{Guowei Zhu}, \bibinfo{person}{Jing Huang}, \bibinfo{person}{Jinfeng Lu}, \bibinfo{person}{Yingyu Luo}, {and} \bibinfo{person}{Tingyu Zhu}.} \bibinfo{year}{2024}\natexlab{}.
\newblock \showarticletitle{Gig to the left, algorithms to the right: A case study of the dark sides in the gig economy}.
\newblock \bibinfo{journal}{\emph{Technological Forecasting and Social Change}}  \bibinfo{volume}{199} (\bibinfo{year}{2024}), \bibinfo{pages}{123018}.
\newblock


\end{thebibliography}

\clearpage
\appendix
\section{Appendix}
\label{sec:appendix}
\subsection{Clustering Results Using Simulated Drivers From Our Algorithms}
\begin{figure}[h]
  \centering
  \includegraphics[width=\linewidth]{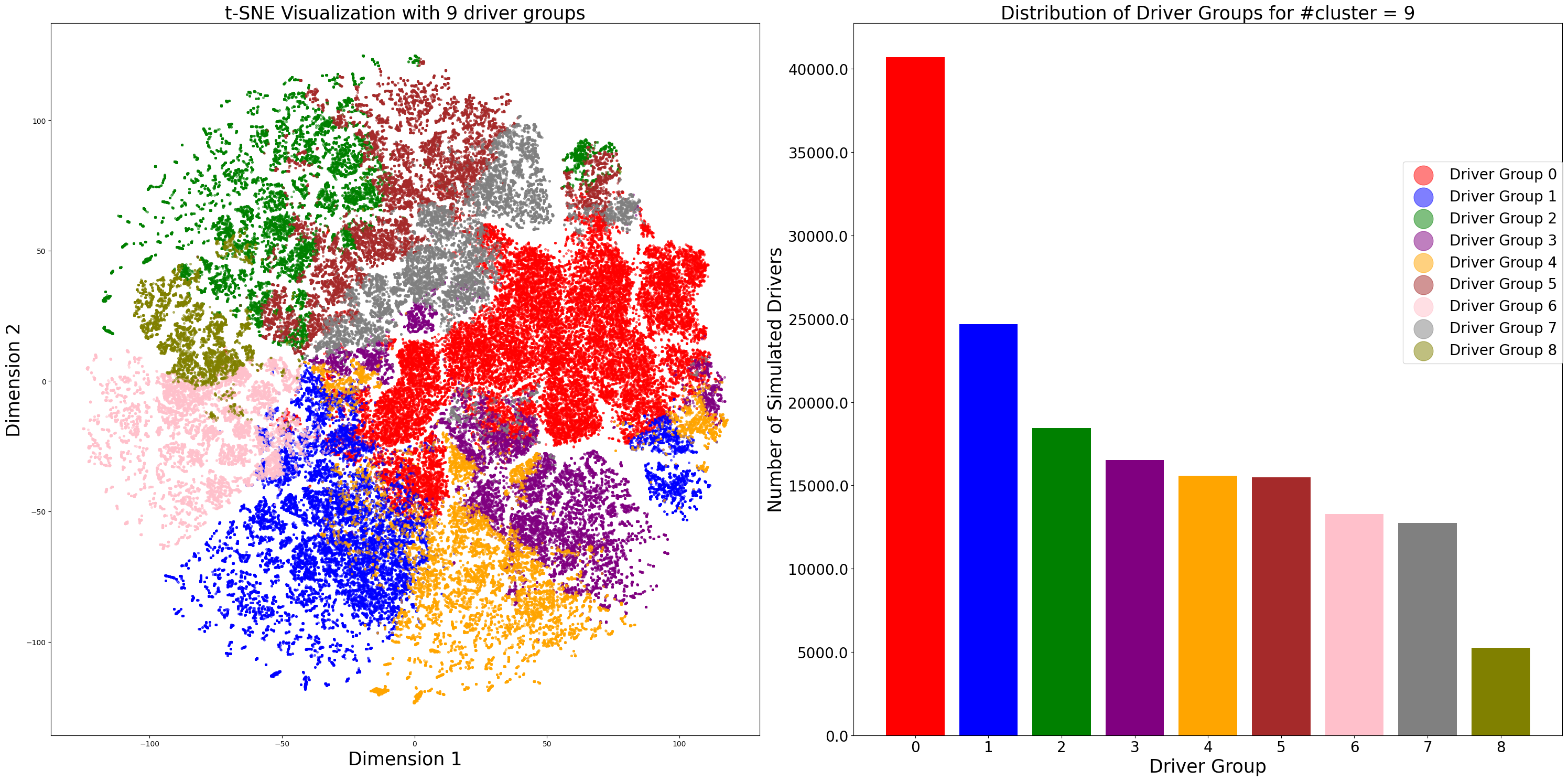}
  \caption{(Left) t-SNE visualization of the best number of clusters using KMeans algorithms and Distribution of clusters on simulated drivers for the week of 2019-08-05. (Right) Distribution of different clusters}
  \label{figure:cluster_2019}
\end{figure}

\begin{figure}[h]
  \centering
  \includegraphics[width=\linewidth]{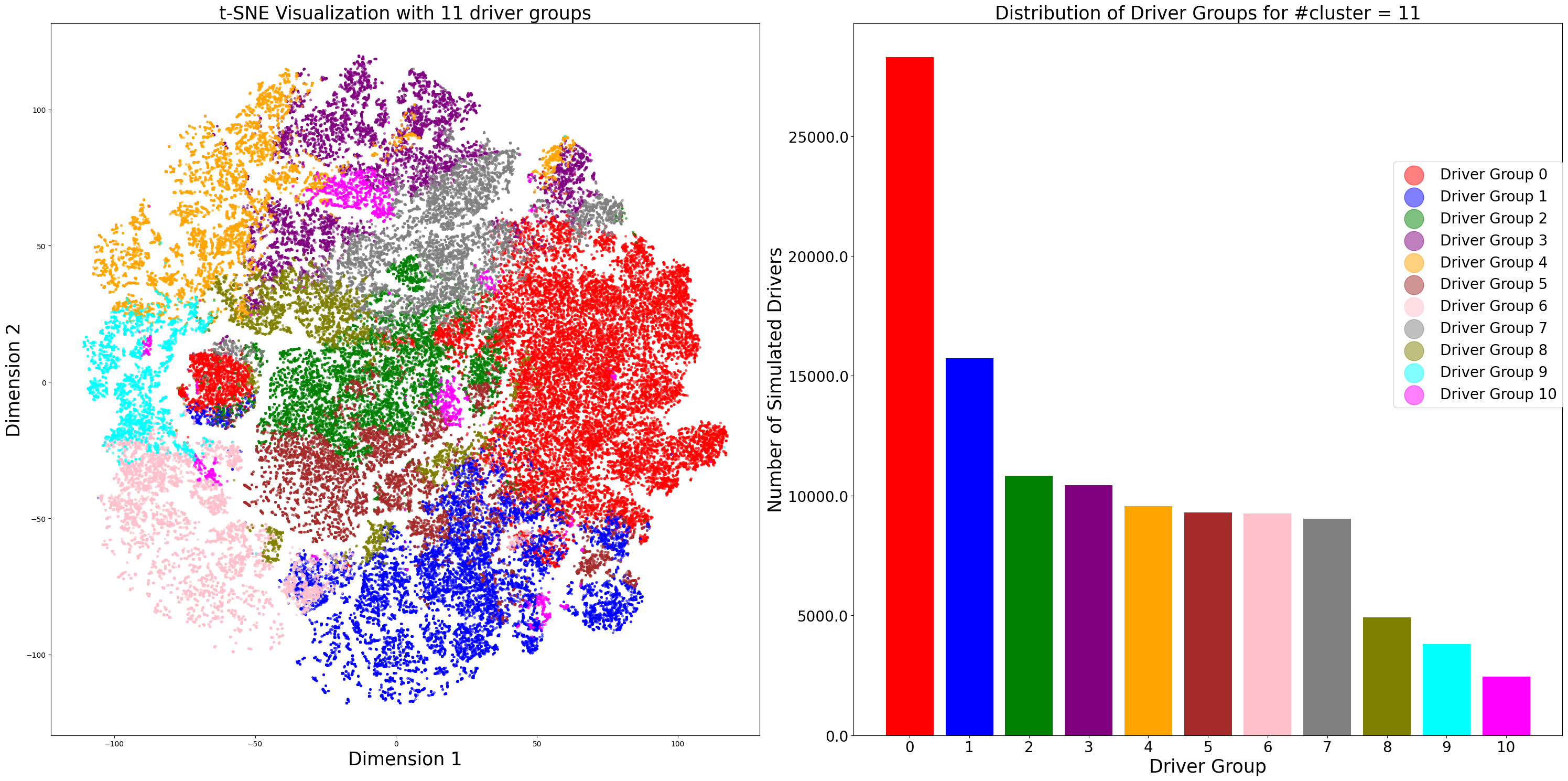}
  \caption{(Left) t-SNE visualization of the best number of clusters using KMeans algorithms and Distribution of clusters on simulated drivers for the week of 2023-08-07. (Right) Distribution of different clusters}
  \label{figure:cluster_2023}
\end{figure}

Figures~\ref{figure:cluster_2019} and~\ref{figure:cluster_2023} illustrate the differences in driver behavior patterns using t-SNE plots, which visually group drivers into clusters. Additionally, we show the distribution of simulated drivers across these clusters. For the week of 2019-08-05, Driver Group 0 accounts for the largest proportion, with over 40,000 simulated drivers. A similar pattern is observed in the week of 2023-08-07, where Driver Group 0 comprises approximately 30,000 simulated drivers.

Additional interesting finding is the emergence of Driver Groups 8, and 10 in the 2023 dataset. These groups represent the smallest number of simulated drivers compared to other clusters and exhibit distinct regional driving patterns that differ significantly from their counterparts in the 2019 data.

This analysis demonstrates the effectiveness of our approach in identifying shifts in driver behavior over time, where the appearance of new driver clusters and their unique characteristics, cannot be captured using the original Chicago dataset, underscoring the value of simulation-based methods for reveal driver behavior's temporal and spatial shifts.

\subsection{Regional Trip Distribution Between 2019-2023}
In this section, we present a comparison of trip distributions across regions from 2019 to 2023. The Central, West Side, and North Side consistently emerge as hotspot areas, collectively accounting for approximately 60--65\% of total trips. Similarly, the least active regions, including the Far Southwest Side and Northwest Side of Chicago, show comparable patterns, contributing only about 2--3\% of total trips.
\begin{figure}[h]
  \centering
\label{distribution_2019}
  \includegraphics[width=\linewidth]{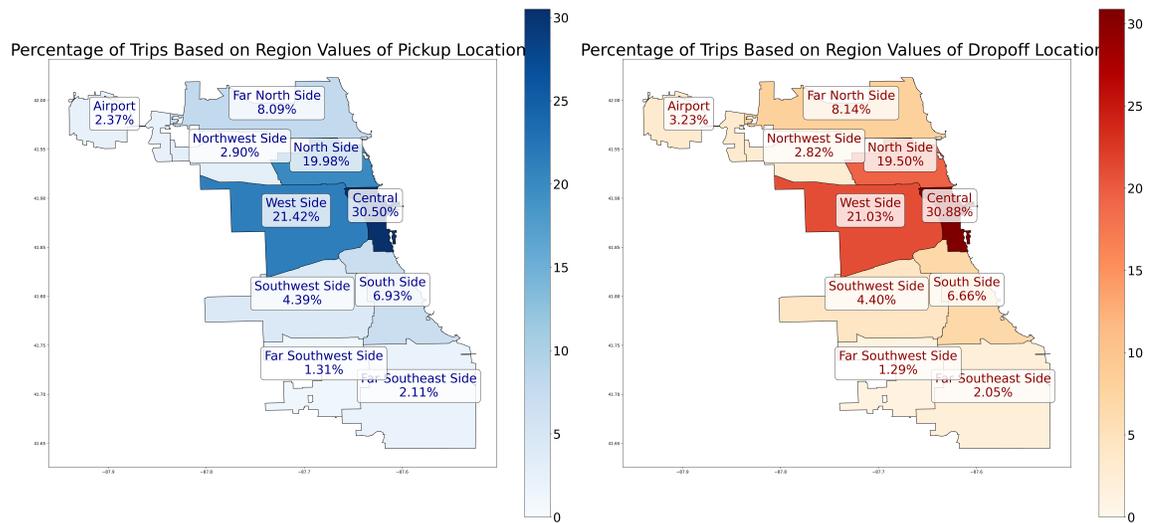}
\caption{Heat maps of pickup and dropoff trip distributions across Chicago regions in 2019.}
\end{figure}

\begin{figure}[h]
  \centering
\label{distribution_2020}
  \includegraphics[width=\linewidth]{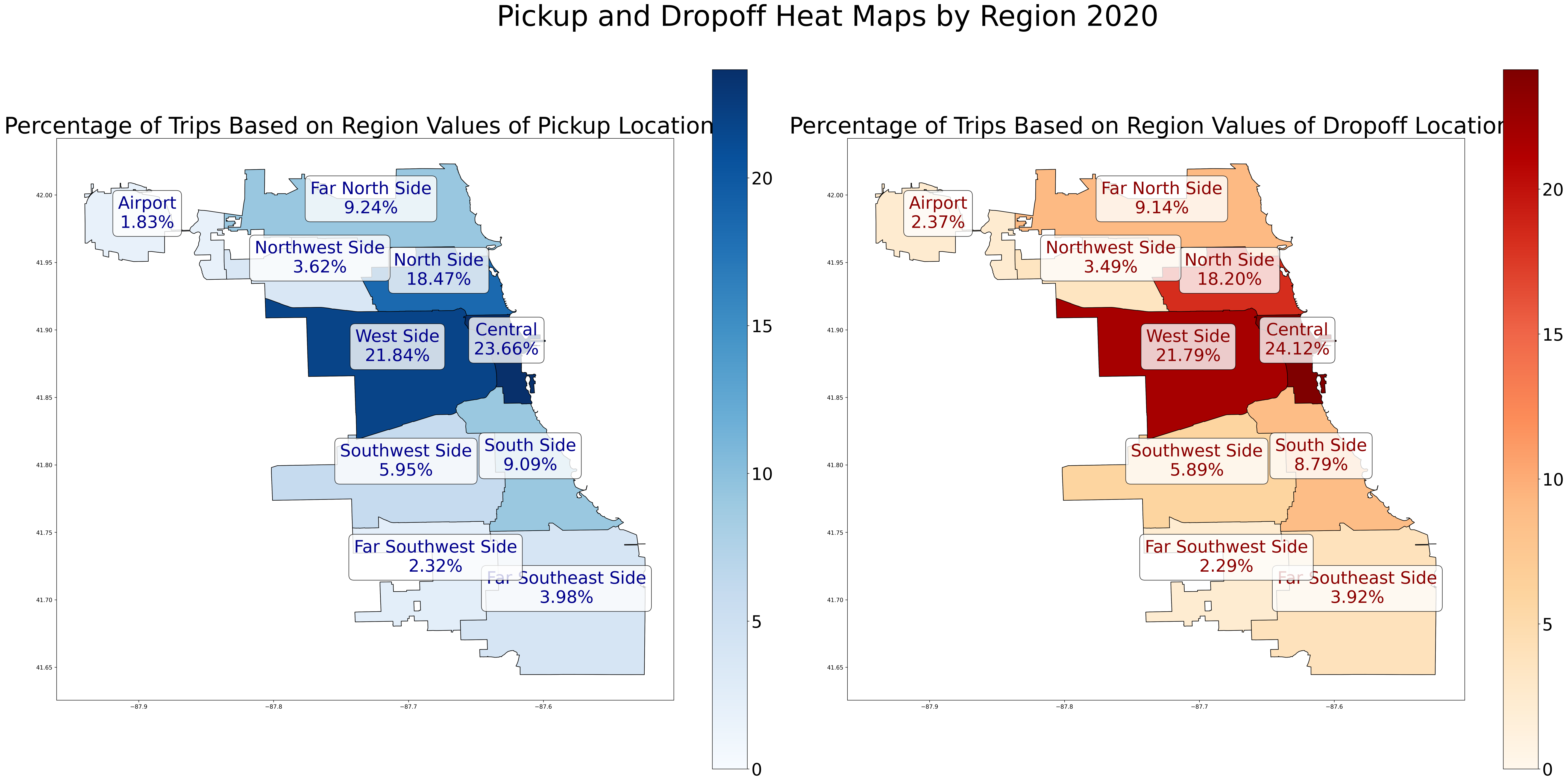}
\caption{Heat maps of pickup and dropoff trip distributions across Chicago regions in 2020.}
\end{figure}

\begin{figure}[h]
  \centering
\label{distribution_2021}
  \includegraphics[width=\linewidth]{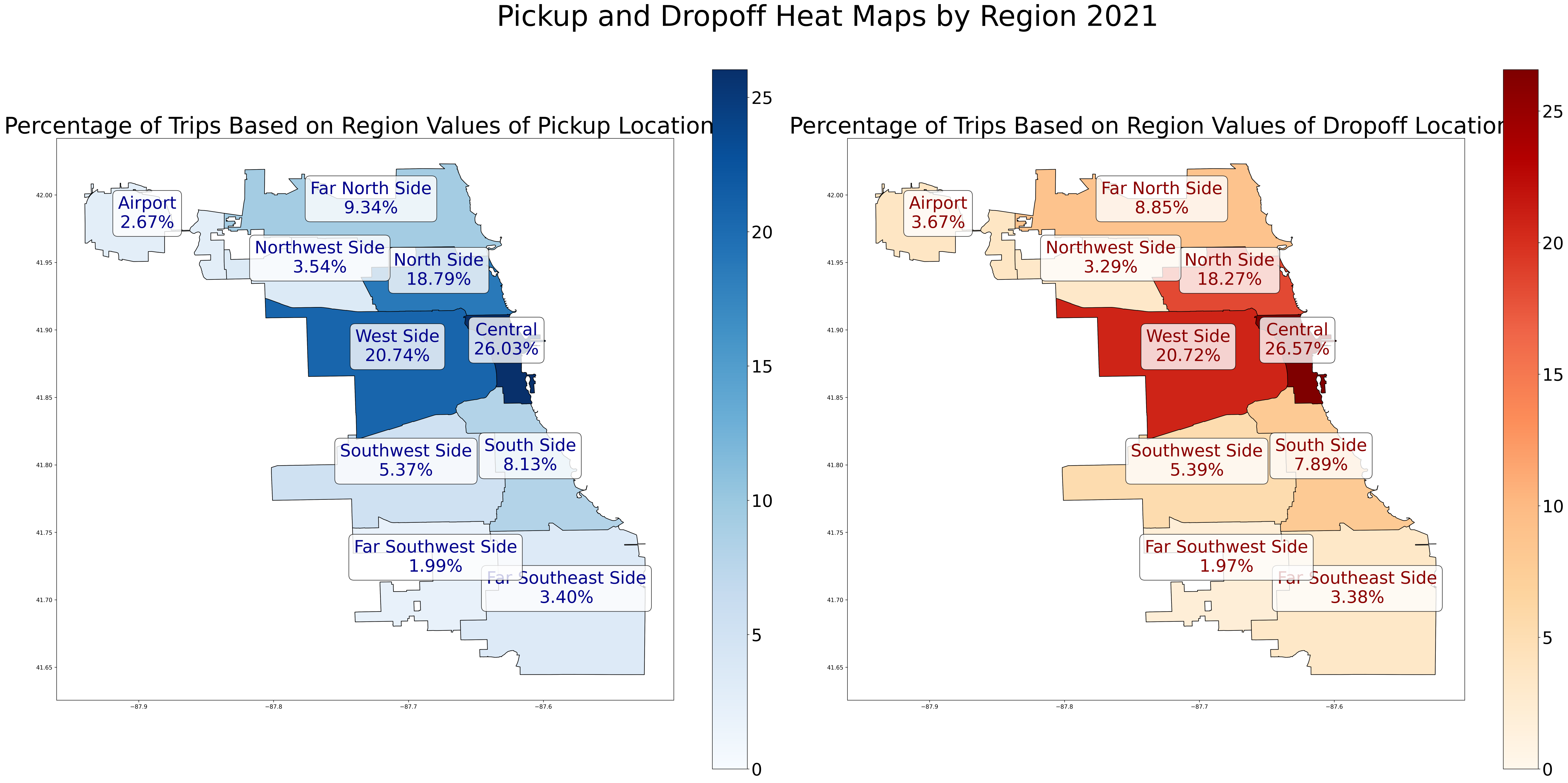}
\caption{Heat maps of pickup and dropoff trip distributions across Chicago regions in 2021.}
\end{figure}

\begin{figure}[h]
  \centering
\label{distribution_2022}
  \includegraphics[width=\linewidth]{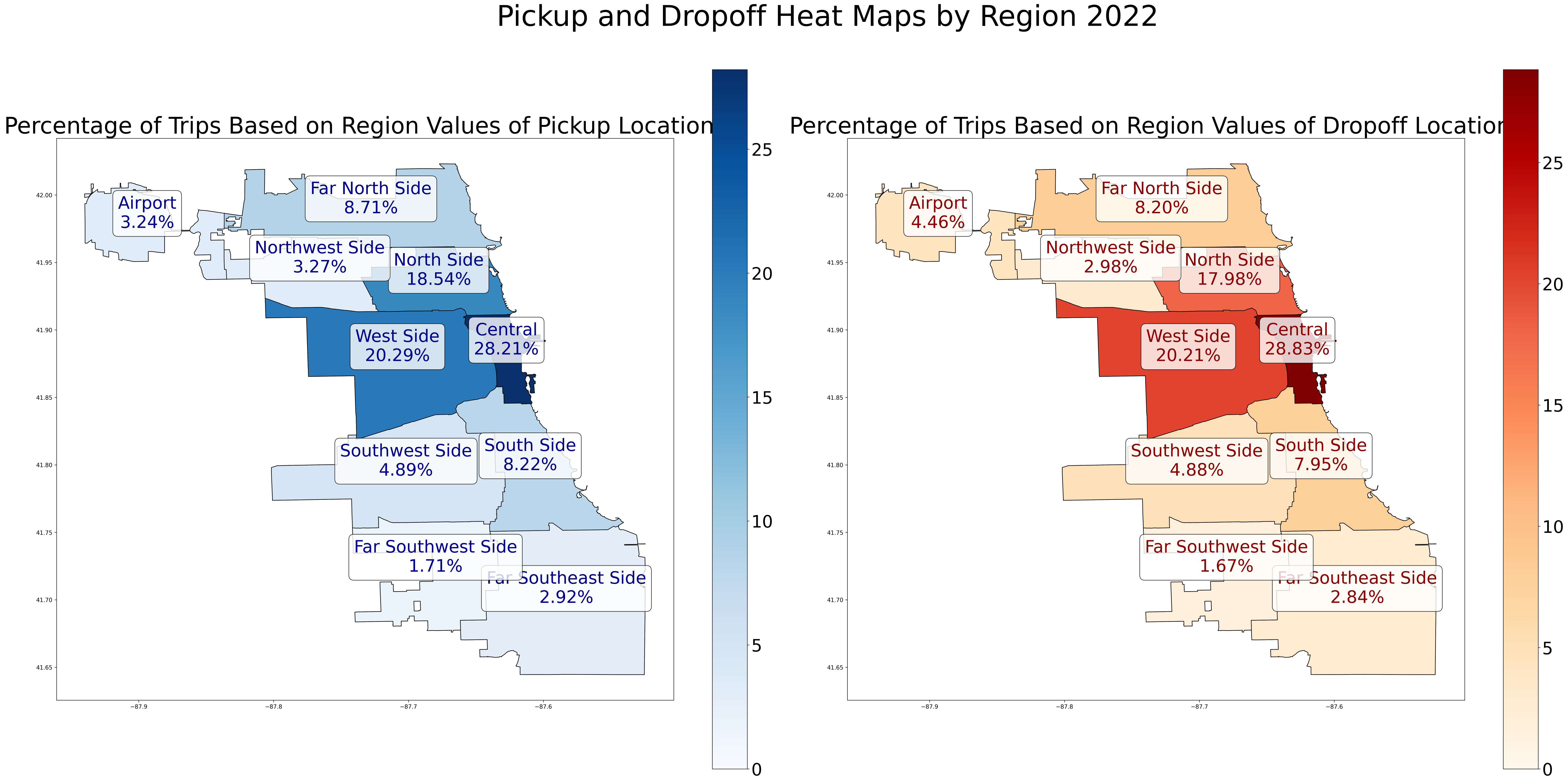}
\caption{Heat maps of pickup and dropoff trip distributions across Chicago regions in 2022.}
\end{figure}

\begin{figure}[h]
  \centering
\label{distribution_2023}
  \includegraphics[width=\linewidth]{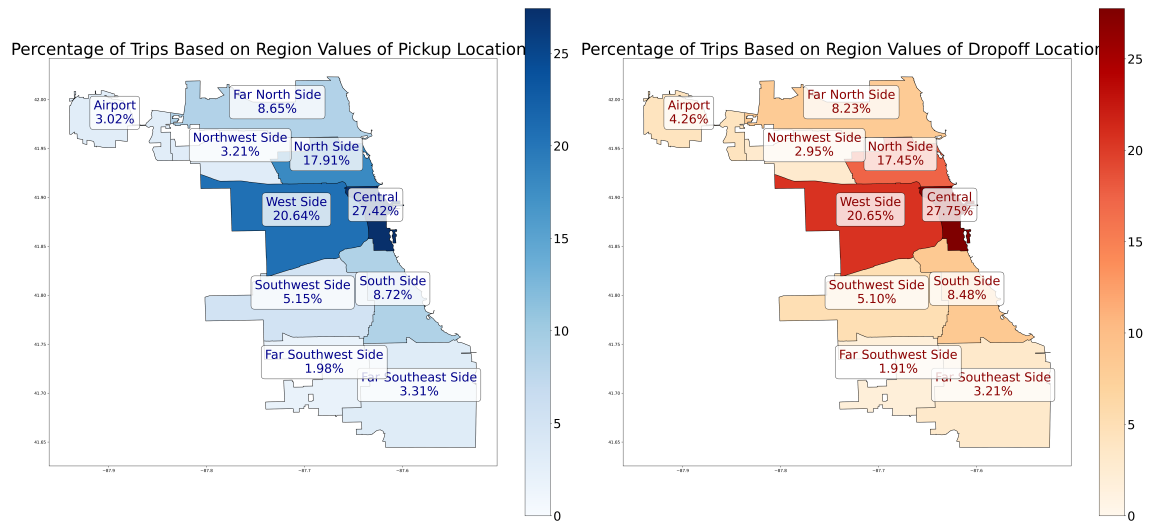}
\caption{Heat maps of pickup and dropoff trip distributions across Chicago regions in 2023.}
\end{figure}

\newpage
\subsection{Regional Trip Cost Between 2019-2023}
In this section, we compare trip costs across regions from 2019 to 2023. While trip distributions remain consistent across years, we observe significant shifts in trip price, aligning with our earlier analysis. These include: (1) an increase in trip costs beginning in 2021, and (2) a substantial pricing gap in the Far South Side areas (Far Southeast Side, Far Southwest Side) compared to top-earning regions like the Central area in the most recent years (2022 and 2023).
\begin{figure}[h]
  \centering
\label{earning_2019}
  \includegraphics[width=\linewidth]{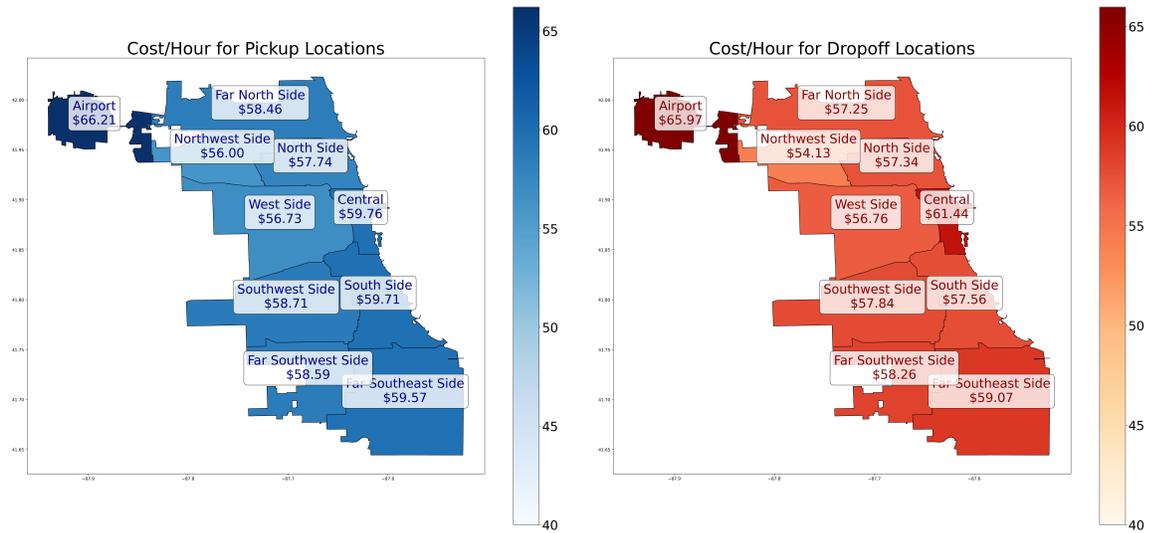}
\caption{Heat maps of pickup and dropoff trip costs across Chicago regions in 2019.}
\end{figure}

\begin{figure}[h]
  \centering
\label{earning_2020}
  \includegraphics[width=\linewidth]{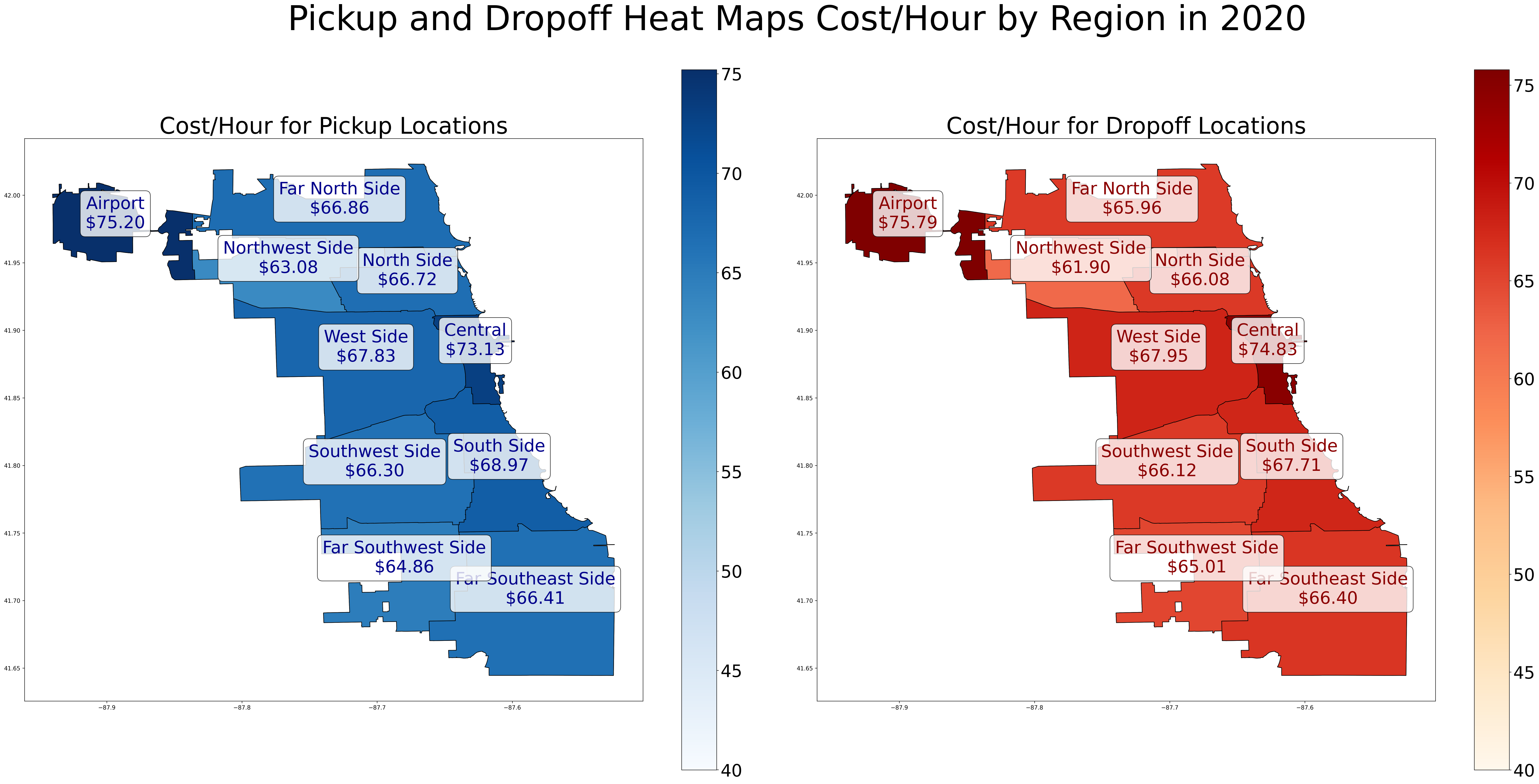}
\caption{Heat maps of pickup and dropoff trip costs across Chicago regions in 2020.}
\end{figure}

\begin{figure}[h]
  \centering
\label{earning_2021}
  \includegraphics[width=\linewidth]{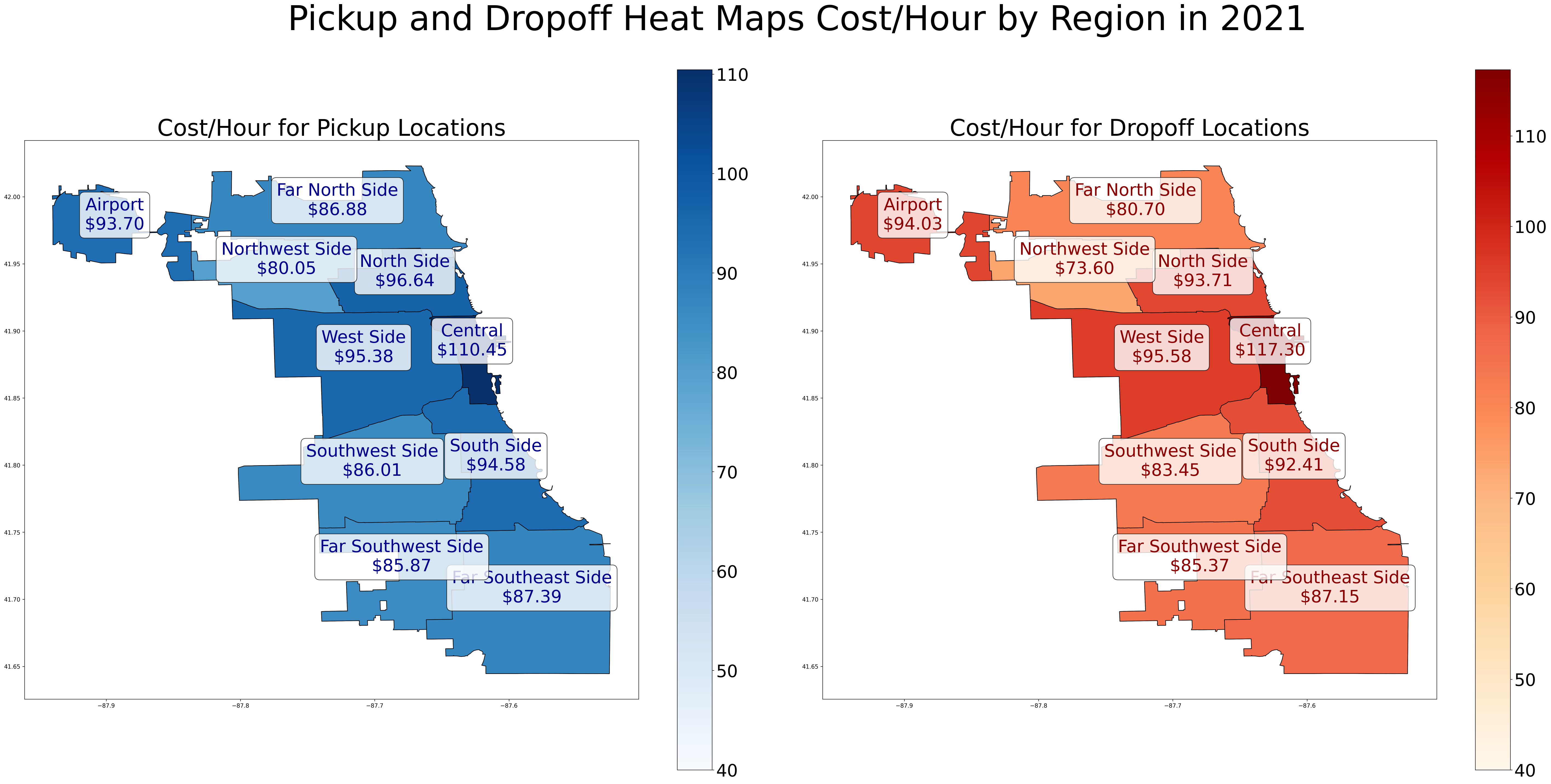}
\caption{Heat maps of pickup and dropoff trip costs across Chicago regions in 2021.}
\end{figure}

\begin{figure}[h]
  \centering
\label{earning_2022}
  \includegraphics[width=\linewidth]{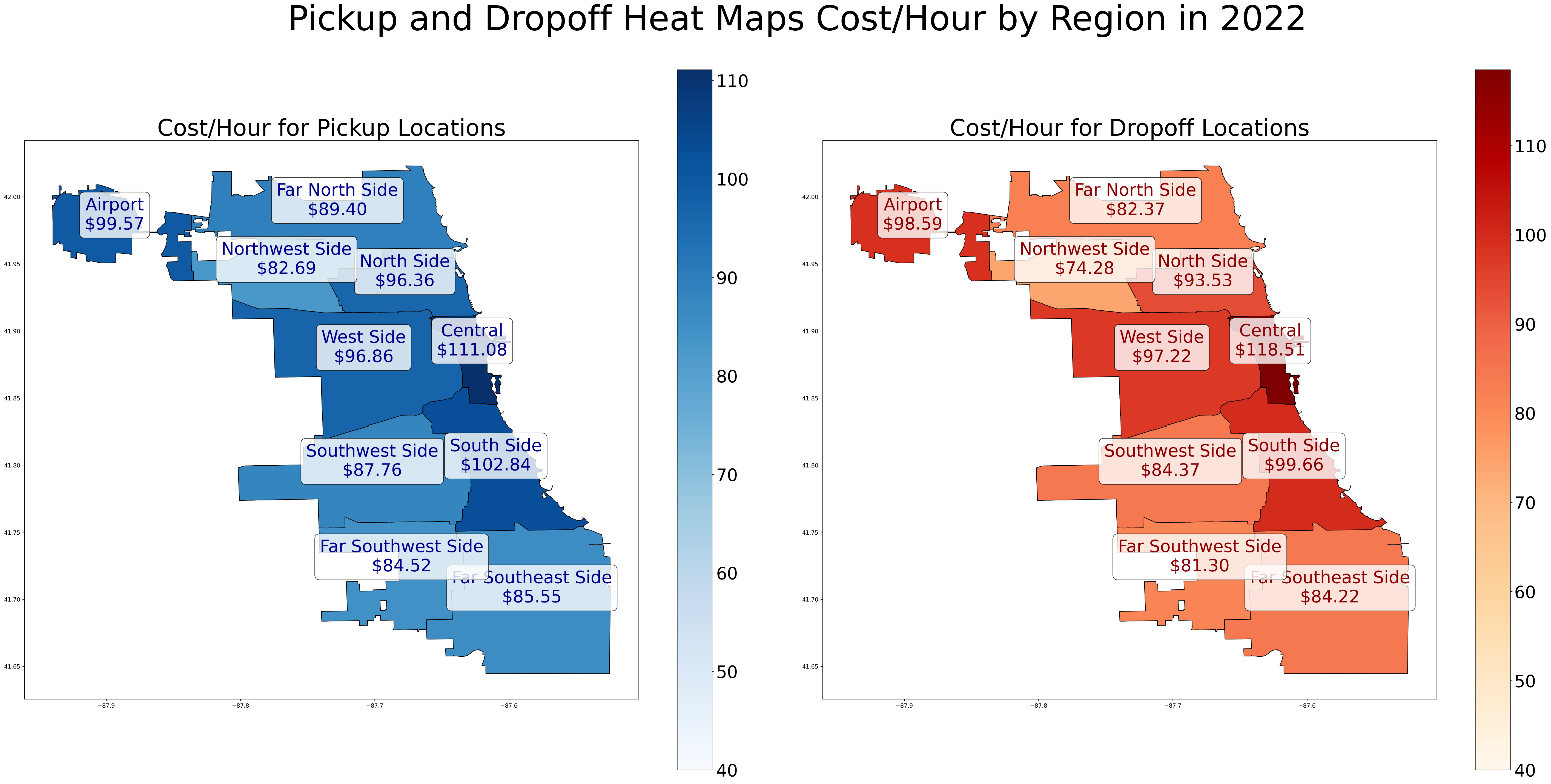}
\caption{Heat maps of pickup and dropoff trip costs across Chicago regions in 2022.}
\end{figure}

\begin{figure}[h]
  \centering
\label{distribution_2023}
  \includegraphics[width=\linewidth]{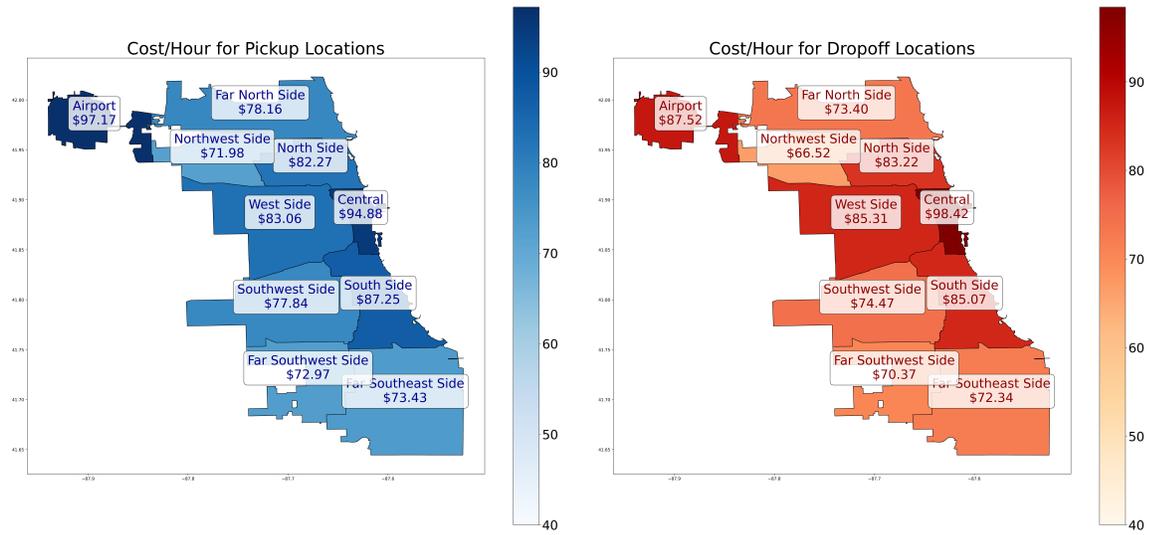}
\caption{Heat maps of pickup and dropoff trip costs across Chicago regions in 2023.}
\end{figure}

\end{document}